\documentclass{article}

\usepackage[preprint]{neurips_2026}

\usepackage[utf8]{inputenc}
\usepackage[T1]{fontenc}
\usepackage{hyperref}
\usepackage{url}
\usepackage{booktabs}
\usepackage{multirow}
\usepackage{amsfonts}
\usepackage{amssymb}
\usepackage{nicefrac}
\usepackage{microtype}
\usepackage{graphicx}
\usepackage{tikz}
\usepackage[version=3]{mhchem}

\newcounter{matbox}
\renewcommand{\thematbox}{\arabic{matbox}}

\title{MatClaw: An Autonomous Code-First LLM Agent for End-to-End Materials Exploration}

\author{
  Chenmu Zhang \\
  Department of Materials Science and NanoEngineering \\
  Rice University \\
  Houston, TX 77005 \\
  \texttt{zcmben2014@gmail.com}
  \And
  Boris I. Yakobson \\
  Department of Materials Science and NanoEngineering \\
  Department of Chemistry \\
  Rice University \\
  Houston, TX 77005 \\
  \texttt{biy@rice.edu}
}

\begin{document}

\maketitle

\begin{abstract}
Existing LLM agents for computational materials science are constrained by pipeline-bounded architectures tied to specific simulation codes and by dependence on manually written tool functions that grow with task scope.
We present MatClaw, a \emph{code-first} agent that writes and executes Python directly, composing any installed domain library to orchestrate multi-code workflows on remote HPC clusters without predefined tool functions.
To sustain coherent execution across multi-day workflows, MatClaw uses a four-layer memory architecture that prevents progressive context loss, and retrieval-augmented generation over domain source code that raises per-step API-call accuracy to ${\sim}$99\%.
Three end-to-end demonstrations on ferroelectric \ce{CuInP2S6} (machine-learning force field training via active learning, Curie temperature prediction, and heuristic parameter-space search) reveal that the agent handles code generation reliably but struggles with tacit domain knowledge.
The missing knowledge, such as appropriate simulation timescales, equilibration protocols, and sampling strategies, is the kind that researchers accumulate through experience but rarely formalize.
Two lightweight interventions, literature self-learning and expert-specified constraints, bridge these gaps, defining a guided autonomy model in which the researcher provides high-level domain knowledge while the agent handles workflow execution.
Our results demonstrate that the gap between guided and fully autonomous computational materials research is narrower than ever before: LLMs already handle code generation and scientific interpretation reliably, and the rapid improvement in their capabilities will accelerate materials discovery beyond what manual workflows can achieve. All code and benchmarks are open-source.
\end{abstract}

\section{Introduction}
\label{sec:intro}

Large language models (LLMs) have emerged as increasingly capable tools for software engineering and scientific research.
On the coding side, frontier models now resolve more than 80\% of real-world GitHub issues in the SWE-bench benchmark~\citep{Jimenez2024}.
On the scientific reasoning side, LLMs exceed human domain-expert accuracy on graduate-level science questions, reaching 94\% on the GPQA Diamond benchmark~\citep{Rein2024}.
These two capabilities converge in computational materials science, where the computational workflow (from structure construction to job submission to result analysis) is executed through code, making it a natural domain for autonomous LLM agents.

Several materials-science agents have been proposed recently, including systems for
autonomous electrochemistry~\citep{Zheng2025},
specific simulation tasks~\citep{Liu2025},
topological materials discovery~\citep{Zhang2025topo}, and
first-principles VASP simulations~\citep{Xia2025vasp}.
In adjacent fields,
Coscientist~\citep{Boiko2023} and
ChemCrow~\citep{Bran2024} demonstrated autonomous chemical research,
El~Agente~\citep{Zou2025} targeted quantum chemistry, and multi-agent frameworks for atomistic simulations have appeared~\citep{Vriza2026,Ansari2024}.

Despite this progress, existing materials agents share two key limitations:

\begin{enumerate}
\item \textbf{Pipeline-bounded architectures.}
Most agents are confined to a fixed set of supported software and predefined task sequences.
For example, VASPilot~\citep{Liu2025} automates only VASP workflows, and Xia~et~al.~\citep{Xia2025vasp} benchmark their agent on a fixed set of first-principles tasks.
In practice, materials research routinely orchestrates multiple codes in a single workflow (e.g., DFT labeling with VASP, force field training with DeePMD-kit, and molecular dynamics simulations), and a pipeline-bounded agent cannot adapt to such multi-code workflows without substantial re-engineering.

\item \textbf{Tool-call dependence.}
The predominant agent design routes actions through manually written tool functions~\citep{Yao2023,Shinn2023}.
This creates a practical scaling problem: expanding to new domains or software packages requires additional hand-written tool functions, and the development effort grows with the scope of tasks.
For instance, ChemCrow~\citep{Bran2024} integrates 18~expert-designed tools to cover its target chemistry tasks; extending it to a new domain would require designing and validating additional tools.
Moreover, complex workflows requiring conditional branching, iterative loops, and error recovery are difficult to express as sequential tool calls.
\end{enumerate}

We present MatClaw, an autonomous \emph{code-first} agent that addresses both limitations.
Rather than calling predefined tools, MatClaw writes and executes Python directly in a sandboxed environment, composing any installed Python library, including, in our demonstrations,
pymatgen~\citep{Ong2013},
atomate2~\citep{Ganose2025},
jobflow~\citep{Rosen2024}, and
DeePMD-kit~\citep{Zhang2018dpmd}.
To sustain coherent execution across long-horizon workflows on remote HPC clusters, MatClaw introduces several mechanisms beyond the basic agent loop.
A retrievable conversation history allows the agent to recall exact parameters and file paths from earlier steps even after they have been pruned from the context window.
A dynamically reloaded experience log, maintained jointly by human feedback and the agent's own accumulated lessons from failures, persists operational knowledge across sessions.
A database query layer provides direct access to exact numerical results (energies, forces, structures) from completed calculations.
For context management, MatClaw uses hierarchical message truncation with a pre-generated summary for each step, avoiding the additional LLM calls that compaction-based approaches require for context compression.
Finally, our benchmarks show that retrieval-augmented generation with structure-aware code chunking substantially improves the agent's ability to use domain APIs correctly and to self-correct when encountering errors (Section~\ref{sec:benchmarks}).

The remainder of this paper is organized as follows.
Section~\ref{sec:design} describes the system architecture, including the \emph{code-first} execution model, four-layer memory, context management strategy, and RAG pipeline.
Section~\ref{sec:demos} validates MatClaw through three end-to-end demonstrations on \ce{CuInP2S6} (CIPS): machine-learning force field distillation, Curie temperature prediction, and heuristic parameter-space search.
Section~\ref{sec:benchmarks} reports retrieval-augmented generation benchmarks comparing chunking methods, retrieval strategies, and LLM providers.
Section~\ref{sec:conclusion} concludes.

\section{System design}
\label{sec:design}

\begin{figure}[t]
\centering
  \includegraphics[width=\textwidth]{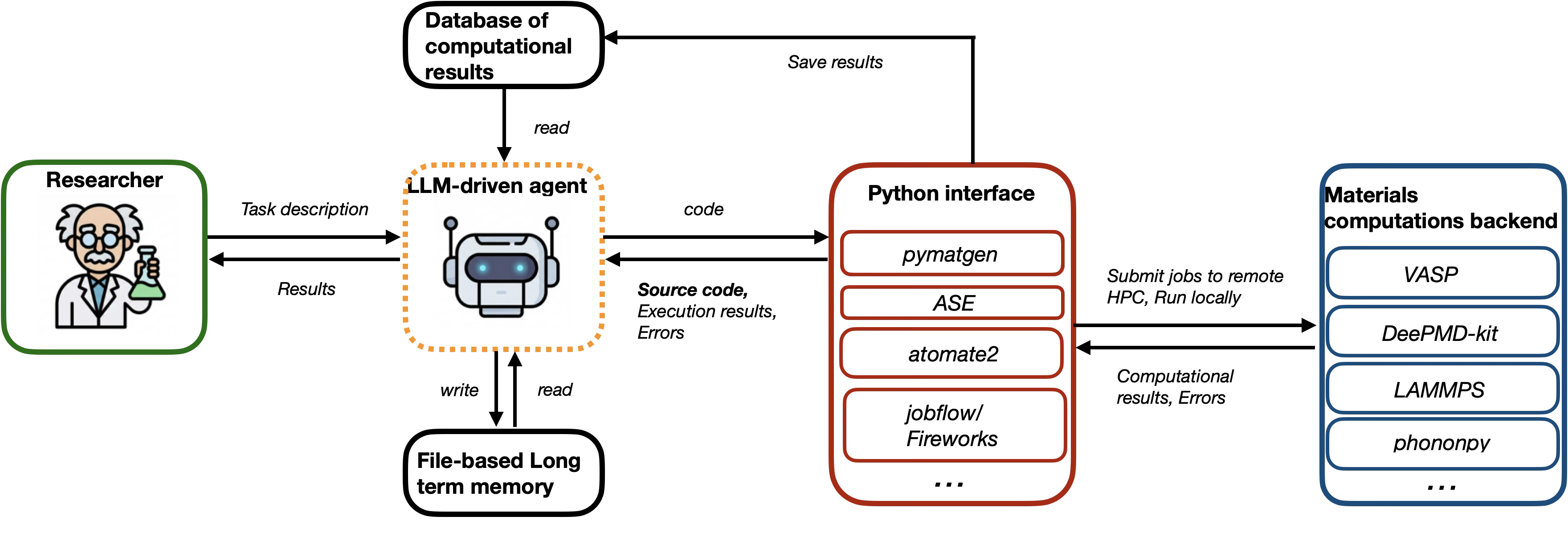}
  \caption{MatClaw architecture. The researcher provides a task description in natural language. The LLM-driven agent generates Python code that composes domain libraries (pymatgen, atomate2, jobflow, etc.), which in turn submit jobs to remote HPC backends (VASP, DeePMD-kit, LAMMPS, etc.) and return computational results. The agent does not directly interact with the backends. File-based long-term memory and a database of computational results provide persistent state across steps and sessions.}
  \label{fig:architecture}
\end{figure}

Figure~\ref{fig:architecture} illustrates the overall architecture.
MatClaw adopts the \emph{code-as-action} paradigm~\citep{Wang2024codeact,Qiao2024taskweaver}: the researcher provides a task description in natural language, and the LLM-driven agent generates Python code that calls domain libraries (pymatgen, ASE, atomate2, jobflow, etc.) through a Python interface.
These libraries in turn submit jobs to materials computation backends (VASP, DeePMD-kit, LAMMPS, phonopy, etc.) on remote HPC clusters or locally, and return computational results and errors back to the agent.
At each step, the agent reads the execution output, reasons about what to do next, and writes new code---adjusting parameters, handling failures, or proceeding to the next stage.
A small set of auxiliary tools handles operations that require blocking waits or remote access, such as HPC job submission and retrieval-augmented search (detailed in Section~\ref{sec:codefirst}).
Two persistence mechanisms support long-running workflows: file-based long-term memory stores operational lessons and conversation history, while a database records exact computational results from completed jobs.
The agent returns final results directly to the researcher.

Two architectural choices distinguish MatClaw from the agents surveyed in Section~\ref{sec:intro}.
First, unlike multi-agent frameworks such as TopoMAS~\citep{Zhang2025topo} and Vriza~et~al.~\citep{Vriza2026}, MatClaw uses a \emph{single agent} that maintains a unified conversation history and memory state, avoiding the coordination overhead and state-synchronization challenges of multi-agent designs.
Second, the agent does not directly manipulate computation backends; instead, it composes Python libraries that handle input generation, job submission, and output parsing.
This indirection means the agent is not coupled to any specific backend and can orchestrate heterogeneous multi-code workflows without re-engineering.

\subsection{Code-first agent design}
\label{sec:codefirst}

In the code-as-action paradigm, the LLM agent's action space is \emph{executable Python code} rather than a fixed set of tool-function calls~\citep{Wang2024codeact,Qiao2024taskweaver}.
Wang~et~al.~\citep{Wang2024codeact} showed that consolidating agent actions into executable code achieves up to 20\% higher success rates compared to JSON- or text-based action formats across 17~LLMs, because code naturally supports composition (calling multiple APIs in sequence), dynamic revision (modifying prior actions based on new observations), and self-debugging through multi-turn interactions with an interpreter.
TaskWeaver~\citep{Qiao2024taskweaver} further demonstrated that a \emph{code-first} framework can incorporate domain-specific knowledge through examples and handle rich data structures that tool-function interfaces struggle to express.

This paradigm is particularly well-suited to computational materials science for two reasons.
First, the domain's mature Python libraries---pymatgen, atomate2, jobflow, dpdata, DeePMD-kit---already encode extensive expert knowledge in the form of input validation, symmetry handling, error recovery, and data format conversion.
By writing Python directly, the agent leverages this embedded expertise for free, rather than duplicating it in hand-written tool wrappers.
Second, the agent can focus on \emph{higher-level scientific reasoning}---deciding what to compute, interpreting results, and planning the next iteration---while the libraries handle low-level details.
A small set of auxiliary tools is retained only for operations that cannot be expressed as local code execution, such as blocking waits for remote HPC jobs and retrieval-augmented search.

In our implementation, each agent step produces a structured response with four fields in a deliberately chosen order: \texttt{phase}, \texttt{plan}, \texttt{code}, and \texttt{summary}.
This ordering exploits autoregressive left-to-right token generation: since each field is conditioned on all preceding fields, the placement order determines what context is available when the LLM generates each part.

\textbf{\texttt{phase}} is generated first.
It is a cross-step topic title that locates the current step within the overall pipeline (\emph{e.g.}, ``Iteration~2, Step~3: Train student ensemble'').
By committing to a pipeline position before elaborating any details, the agent is forced to anchor its state awareness---preventing goal drift when older context has been pruned.
Writing the phase first also forces the LLM to implicitly summarize past progress, reinforcing continuity across steps.

\textbf{\texttt{plan}} comes second, benefiting from the phase context (``Since I am in iteration~2, I need to\ldots'').
The plan describes the specific actions for this step and how they contribute to subsequent steps, acting as a specification that the code must satisfy.

\textbf{\texttt{code}} is generated third, with both the phase and plan already written.
This gives the code generation maximum context for correctness---the plan serves as a natural-language contract that the executable code implements.

\textbf{\texttt{summary}} is generated last, after the code field.
Because the summary is conditioned on the concrete code the model has already written, it reflects the generated action rather than only the earlier planning intent.
This distinction is critical: summaries are persisted to disk and serve as the primary index for conversation history retrieval (Section~\ref{sec:memory}).
If the summary were placed before the code, it would describe intent that may diverge from the actual implementation (due to error handling, fallback paths, or additional operations added during code generation), creating inaccurate history indices.
The summary field thus bridges the \emph{code-first} execution model and the memory architecture: it provides a zero-cost, per-step annotation that enables context compression without additional LLM calls (Section~\ref{sec:context}).

\subsection{Four-layer memory architecture}
\label{sec:memory}

Long-running agent workflows are susceptible to what we term the \textbf{Sisyphus Trap}---a progression of three failure modes that emerge as conversations grow beyond the LLM's context window:

\begin{itemize}
\item \emph{Detail loss}: The agent forgets exact file paths and parameter values, wasting steps re-discovering information already established earlier.
\item \emph{Goal drift}: As context pruning removes early planning messages, the agent loses track of iteration counts, convergence criteria, and simulation settings, causing it to repeat or contradict earlier decisions.
\item \emph{Catastrophic forgetting}: After multiple pruning cycles, the agent loses all memory of its own pipeline and may attempt to restart from scratch, wasting all prior computation.
\end{itemize}

These failure modes are consistent with our observations in multi-day workflows and echo challenges identified in prior work.
Packer~et~al.~\citep{Packer2024memgpt} showed that limited context windows severely degrade performance in extended conversations and document analysis, and proposed MemGPT, an OS-inspired system that pages data between the LLM context (``main memory'') and external storage.
Wang~et~al.~\citep{Wang2023voyager} demonstrated that a persistent skill library alleviates catastrophic forgetting in embodied agents by preserving learned behaviors across sessions.
More broadly, the CoALA framework~\citep{Sumers2024coala} organizes agent memory into \emph{working memory} (active information for the current decision cycle), \emph{episodic memory} (past experiences), \emph{semantic memory} (world knowledge), and \emph{procedural memory} (skills and code), providing a principled taxonomy for these mechanisms.

Drawing on these ideas, MatClaw implements four memory layers with complementary storage and access patterns (Table~\ref{tbl:memory}), each mapped to CoALA's taxonomy.

\begin{table}[t]
  \caption{Four-layer memory architecture mapped to the CoALA taxonomy~\citep{Sumers2024coala}. Each layer addresses a different recall problem.}
  \label{tbl:memory}
  \centering
  \small
  \begin{tabular}{lllll}
    \toprule
    Layer & CoALA category & Purpose & Storage & Access \\
    \midrule
    1. In-context & Working memory & Active state & LLM context & Always in context \\
    2. History & Episodic memory & Recall pruned steps & Append-only file & On-demand retrieval \\
    3. Experience & Semantic memory & Cross-session lessons & Editable text file & Always in context \\
    4. Database & External grounding & Numerical results & Job-store database & On-demand retrieval \\
    \bottomrule
  \end{tabular}
\end{table}

\paragraph{Layer~1: In-context working memory.}
All messages currently in the LLM's context window constitute the agent's working memory in CoALA's sense---the active information available for the current decision cycle.
The structured output fields (Section~\ref{sec:codefirst}) enhance this layer: the \texttt{phase} field of the most recent step always anchors the agent's position in the pipeline, even when older messages have been pruned.
Zone-based context management (Section~\ref{sec:context}) governs what remains in this layer.

\paragraph{Layer~2: Episodic conversation history.}
When context pruning removes earlier messages from working memory, the agent loses access to exact parameter values, file paths, and error messages from those steps.
Layer~2 addresses this by maintaining a persistent, append-only record of every message exchanged during the run---the agent's episodic memory in CoALA's terms.
When the agent needs to recall information from a pruned step, it first scans one-line summaries (produced by the \texttt{summary} field of each step, Section~\ref{sec:codefirst}) to locate the relevant step, then retrieves the full content of that step on demand.
This two-stage retrieval---summary scan followed by selective detail loading---avoids the LLM-based summarization that systems like MemGPT~\citep{Packer2024memgpt} use to compress context, instead relying on the pre-generated summaries that are already available at zero additional cost.

\paragraph{Layer~3: Semantic experience log.}
A persistent text file stores operational lessons that the agent or human operators accumulate across sessions---for example, ``remote job submission requires uploading input files before launching the workflow'' (discovered from a failed HPC run).
In CoALA's terms, this is writable semantic memory: knowledge about the world and the agent's own constraints, incrementally refined through experience.
Unlike Voyager's skill library~\citep{Wang2023voyager}, which stores executable code (procedural memory), MatClaw's experience log stores natural-language rules that are injected into the system prompt.

A distinctive feature is \emph{dynamic reloading}: the experience file is monitored for changes and re-read before each agent step, so lessons added by the agent or edited by the researcher take effect immediately without restarting the agent.

\paragraph{Layer~4: External database.}
A read-only query layer over the workflow job store gives the agent direct access to exact numerical results (energies, forces, structures) from completed calculations.
This corresponds to external grounding in CoALA's action taxonomy---the agent queries a digital environment (the database) rather than relying on conversation history, which may have been pruned.
This layer is essential for steps 40+ where earlier tool outputs have been removed from the context window.

\subsection{Context management}
\label{sec:context}

Recent studies have shown that LLM reasoning quality degrades as input length increases, even well within the advertised context window.
Liu~et~al.~\citep{Liu2024lost} demonstrated that performance drops when relevant information is positioned in the middle of long contexts, and the
``context rot'' phenomenon~\citep{ContextRot2025} shows measurable degradation simply from increasing input length, regardless of content relevance.
For long-running agents, these effects compound: every additional step adds messages to the context, steadily eroding reasoning quality.
MatClaw therefore caps the effective context window conservatively per provider---for example, 200K tokens for models that advertise 1M---informed by these degradation studies and our internal multi-needle retrieval tests.

When the context exceeds this cap, the agent must compress its history.
A common approach is \emph{LLM-based compaction}, in which an additional LLM call summarizes old messages before discarding them~\citep{Kang2025acon}.
This is effective but costly: the summarization call itself consumes tokens, and the generated summary may lose important details.
Recent work by Lindenbauer~et~al.~\citep{Lindenbauer2025} showed that a simpler strategy---\emph{observation masking}, which replaces old tool outputs with placeholders while preserving the agent's reasoning trace---halves cost while matching LLM summarization's task-completion rate on the SWE-bench benchmark.

MatClaw builds on this finding with a \textbf{zone-based pruning} scheme that applies progressively aggressive compression from newest to oldest messages.
When total tokens exceed the context cap, four zones receive different treatment:
the newest messages are fully protected;
the next tier has tool responses trimmed to a head-and-tail excerpt;
an older tier replaces tool responses with short placeholders (analogous to the observation masking of Lindenbauer~et~al.);
and the oldest messages are removed entirely, replaced by a single truncation marker.
Bootstrap messages (system prompt and initial task description) are always protected regardless of zone, consistent with the attention-sink phenomenon identified by Xiao~et~al.~\citep{Xiao2024streamingllm}, which showed that initial tokens play a disproportionate role in maintaining LLM output stability.

A key advantage of this design is that \emph{no information is permanently lost}.
The full conversation history is retained on disk in the episodic store (Layer~2), and the pre-generated \texttt{summary} field from each step (Section~\ref{sec:codefirst}) serves as a lightweight index for on-demand recovery.
When the agent needs to recall pruned information, it scans the summary index to locate the relevant step and retrieves the full content---without requiring an additional LLM summarization call.

\subsection{Retrieval-augmented generation}

For a \emph{code-first} agent, correct API usage is the most fundamental requirement: every step generates Python code that calls domain library functions, and a single incorrect function name, wrong parameter, or outdated import path causes the step to fail.
In our benchmarks (Section~\ref{sec:benchmarks}), the per-question error rate without RAG ranges from 5\% to 30\% depending on the task category, and because each workflow step compounds these errors, overall workflow reliability degrades rapidly as the number of API calls grows.
Retrieval-augmented generation (RAG) over domain source code is therefore essential for sustained correctness.

To this end, we invested substantial effort in code-aware retrieval.
Our approach uses \emph{context enrichment}---retrieving verbatim source code snippets and injecting them into the agent's prompt---built on structure-aware code chunking methods~\citep{Zhang2025cast,codechunk} that split source files at AST boundaries to produce semantically coherent units.
Retrieval uses BM25 with three-query reciprocal rank fusion~\citep{Cormack2009} to improve recall.
We compared three chunking methods on a code QA benchmark and two retrieval strategies on a documentation QA benchmark (Section~\ref{sec:benchmarks}), finding that structure-aware chunking with BM25 retrieval yields the highest end-to-end accuracy.

\section{Demonstration tasks and failure analysis}
\label{sec:demos}

We validate MatClaw on a research project involving molecular dynamics (MD) simulations of monolayer \ce{CuInP2S6} (CIPS).
CIPS is a van der Waals layered ferroelectric in which Cu atoms occupy a double-well potential along the out-of-plane direction; an applied electric field can drive Cu atoms from one potential well to the other, switching the local polarization.
He~et~al.~\citep{He2023} trained a DeePMD potential for CIPS using the DP-GEN concurrent learning scheme~\citep{Zhang2020dpgen} and used it to simulate polarization switching under electric fields at elevated temperatures, observing spatially uncorrelated switching in which each Cu site nucleates independently---consistent with the nucleation-limited switching (NLS) picture.
In conventional ferroelectrics, however, polarization reversal can also proceed through field-driven domain wall motion: walls behave as pinned elastic interfaces that advance by thermally activated creep at low drive and cross over to depinning at higher fields~\citep{Paruch2013,Liu2016nature}.
This raises the question of whether CIPS, despite the NLS-like behavior reported at elevated temperatures, can access a more correlated wall-propagation regime under different conditions---specifically, whether lowering temperature to suppress stochastic thermal nucleation while increasing the electric field can drive a crossover from independent local flipping to sequential domain wall propagation.
The boundary between these regimes in electric field--temperature parameter space has not been mapped.

To investigate this, we would need three things: (1)~a machine-learning potential validated across a wide range of thermodynamic conditions, including the ferroelectric switching barrier; (2)~knowledge of the model's Curie temperature $T_\mathrm{c}$, to distinguish thermal from field-driven behavior; and (3)~a systematic exploration of the $(E, T)$ parameter space to locate the domain wall propagation regime.
We therefore decompose this research goal into three agent tasks, each testing a different aspect of autonomy (Table~\ref{tbl:demos}):
(a)~training a machine-learning force field via active learning,
(b)~determining the Curie temperature from MD simulations, and
(c)~heuristic search through $(E, T)$ space for domain wall propagation, rather than exhaustive grid sweeping.

\begin{table}[t]
  \caption{Summary of demonstration runs on monolayer \ce{CuInP2S6}. Tasks~1 and~2 were each attempted twice; the first attempt reveals a characteristic failure mode, and the second incorporates a targeted intervention.}
  \label{tbl:demos}
  \centering
  \small
  \begin{tabular}{llclll}
    \toprule
    Task & Run & Steps & LLM & Intervention & Outcome \\
    \midrule
    \multirow{2}{*}{1. MLP distillation}
      & 1a & 21 & Opus~4.6 & None & Failure \\
      & 1b & 45 & Opus~4.6 & Paper + constraint & Success \\
    \addlinespace
    \multirow{2}{*}{2. Curie $T_\mathrm{c}$ exploration}
      & 2a & 17 & GPT-5.4 & None & Failure \\
      & 2b & 8  & GPT-5.4 & Convergence check & Success \\
    \addlinespace
    3. $(E, T)$ search
      & 3  & 27 & Opus~4.6 & Metric provided & Success \\
    \bottomrule
  \end{tabular}
\end{table}

All three tasks use the pre-trained DeePMD model from He~et~al.~\citep{He2023} as the ground-truth potential rather than DFT calculations, reducing labeling cost from hours per frame (VASP) to seconds (teacher model inference).
This substitution does not affect the test of agent capability: the workflow orchestration---supercell construction, MD submission, data conversion, active learning iteration, and convergence checking---is identical regardless of whether labels come from DFT or a teacher model.
Tasks~2 and~3 use the teacher model directly for production MD.

\subsection{Task 1: ML force field distillation via active learning}
\label{sec:demo1}

Training a machine-learning interatomic potential via active learning is a multi-stage iterative workflow: train an ensemble of models, explore configuration space with MD, identify high-uncertainty structures through inter-model disagreement, label them with a reference method, and retrain~\citep{Zhang2020dpgen}.
The central challenge is not the workflow orchestration itself, but the design of adequate sampling strategies---choosing simulation conditions (temperatures, pressures, simulation lengths) that cover the relevant physics, such as phase transitions and barrier crossings, rather than merely sampling near-equilibrium harmonic oscillations.

The task description given to the agent (abridged) reads:

\begin{quote}
\small
\textit{I have a monolayer CuInP$_2$S$_6$ (CIPS) structure in `./CuInP2S6.cif' (10 atoms/cell). [\ldots] Train a ``student'' DeePMD model for CIPS by distilling from a published teacher model. [\ldots] Use the teacher model to generate initial training data by running MLFF MD at diverse thermodynamic conditions, then train student models on the teacher's predictions. Iterate with active distillation: explore new structures with student model MD, evaluate by different student models, select high-variance frames from student models, label with the teacher model, and retrain the students.}

\textit{STOPPING CONDITION: Stop when EITHER: (a) You have completed 5 active distillation iterations, OR (b) The student model achieves MAE\_f below 0.10~eV/A on a held-out test set that was never used for training.}

\textit{Constraints: Type map: Cu, In, P, S. Keep supercells $<$ 200 atoms. Use the `fast' DeePMD network preset. [\ldots] Break your work into phases. After each phase completes, inspect the results before proceeding.}
\end{quote}

The prompt constrains the workflow structure (supercell size, phased execution, output format) but deliberately leaves the sampling strategy open: the agent must decide what temperatures to simulate, how long to run each MD trajectory, and how to select frames for retraining.
This design tests whether the agent can make sound experimental design choices---the kind of tacit knowledge that distinguishes an expert from a novice---in addition to executing the workflow mechanics.

\paragraph{First attempt: inadequate configuration-space sampling.}
The agent completed the workflow in 21~steps over 44~minutes:
\begin{itemize}
\item \textbf{Phase~1} (steps~1--7): Loaded the structure, searched the documentation for the force-field MD API, and resolved an incorrect enum value through iterative RAG search.
\item \textbf{Phase~2} (steps~8--10): Submitted five teacher-model MD jobs at 200--500~K and waited for completion.
\item \textbf{Phase~3} (steps~11--17): Encountered the MongoDB document-size limit when passing 102~MB of trajectory data inline, and autonomously recovered by converting to the DeePMD numpy format and uploading to the remote cluster.
\item \textbf{Phase~4} (steps~18--21): Trained two student models with different random seeds, evaluated them on a held-out test set at 350~K, and reported MAE$_\mathrm{f} = 0.082$~eV/\AA---meeting the stopping condition at iteration~0.
\end{itemize}
Although all steps executed correctly and both errors were self-corrected, the resulting model is unreliable for ferroelectric simulations.
The agent chose 1~ps MD trajectories (500 steps $\times$ 2~fs), which sample only harmonic oscillations near the equilibrium Cu position; no Cu atom crosses the ferroelectric switching barrier at any temperature.
The held-out test set (350~K, 0.6~ps) is drawn from the same narrow distribution, so the low MAE$_\mathrm{f}$ reflects interpolation within a small region of configuration space rather than genuine generalization.
A domain expert would know that Cu barrier crossing in CIPS requires trajectories of $\gtrsim$10~ps, and that a test set from the same short-trajectory distribution does not validate the model for the switching dynamics that matter.
This gap---choosing appropriate simulation timescales for the physics of interest---is precisely the kind of tacit knowledge that the agent lacks.

\paragraph{Second attempt: literature-guided active learning.}
To address this, we modified the task description in two ways.
First, we prepended a ``Background reading'' section that provides the He~et~al.\ paper~\citep{He2023} and instructs the agent to extract the active learning methodology and persist key points as experience notes before proceeding with the task.
Second, we added a single constraint: ``Initial teacher MDs: at least 20~ps per temperature.''
The rest of the prompt---workflow structure, stopping condition, output format---remained identical to the first attempt.
The prepended section reads:

\begin{quote}
\small
\textit{`./He\_paper.pdf' describes the DP-GEN concurrent learning methodology for CuInP$_2$S$_6$. Read it with read\_pdf, extract the active learning strategy (model count, exploration, selection criteria, convergence), and persist the key points as experience notes via write\_experience. Then proceed with the task.}
\end{quote}
The agent completed this run in 45~steps:
\begin{itemize}
\item \textbf{Phase~1} (steps~1--3): Read the paper and extracted the DP-GEN methodology into persistent experience notes (Box~\ref{box:experience}), identifying the key design choices: an ensemble of four models, inter-model force deviation $\sigma$ as the selection criterion, and three selection bands.
The physical reasoning behind these bands is important: $\sigma < 0.05$~eV/\AA\ means all models agree, indicating the configuration is already well-represented in the training data; $0.05 < \sigma < 0.15$~eV/\AA\ means the models disagree enough that the configuration is informative but still physically reasonable; $\sigma > 0.15$~eV/\AA\ means the configuration is so far from the training distribution that the models' predictions are unreliable, often indicating unphysical distortions.
This three-band selection scheme is a standard practice in the DP-GEN community but is not documented in any tutorial or user guide---it is tacit knowledge passed through research papers and personal communication, which is precisely why the agent could not discover it autonomously in the first attempt.

\begin{table}[t]
\centering
\small
\refstepcounter{matbox}\label{box:experience}
\fbox{\parbox{0.92\textwidth}{%
\textbf{Box~\thematbox.} Experience note extracted by the agent from He~et~al.~\citep{He2023}.\\[4pt]
\texttt{Key parameters from He et al. for CuInP2S6 DP model training:}\\
\texttt{- 4 independent DP models (ensemble of 4)}\\
\texttt{- Concurrent learning: train -> explore (NPT MD) -> select -> label -> retrain}\\
\texttt{- Selection criterion: max force deviation $\sigma$}\\
\texttt{\quad $\sigma$ < 0.05 eV/\AA: well described, skip}\\
\texttt{\quad 0.05 < $\sigma$ < 0.15 eV/\AA: label and add to training set}\\
\texttt{\quad $\sigma$ > 0.15 eV/\AA: too distorted, skip}\\
\texttt{- Convergence: all explored configs have $\sigma$ < 0.05}\\
\texttt{- 23 iterations, 11{,}260 training configurations}
}}
\end{table}

\item \textbf{Phase~2} (steps~4--19): Searched the API documentation, submitted four teacher-model MD runs at 100, 300, 500, and 800~K (20~ps each), encountered the same MongoDB size limit and sandbox import restriction as in the first attempt, recovered from both, and trained two student models.
The 20~ps constraint forced the training data to sample Cu barrier crossings: at 500~K, 38\% of Cu atom-frames crossed the midplane, and at 800~K, 45\% crossed.
This made the initial MAE$_\mathrm{f}$ higher (0.103~eV/\AA\ vs.\ 0.082 in the first attempt) because the test set now included the difficult barrier-crossing configurations---which in turn triggered the active learning loop.

\item \textbf{Phase~3} (steps~20--44): Executed two active learning iterations.
In iteration~1, it explored at 200~K with a student model, found all 501 frames in the $[0.05, 0.15]$ selection band, subsampled 200 frames, labeled them with the teacher, and retrained (MAE$_\mathrm{f}$ remained at 0.103).
In iteration~2, it explored at 600~K and 1000~K; at 1000~K, the student model showed extreme force deviations ($\sigma_\mathrm{max} = 87.5$~eV/\AA), with only 26 of 501 frames falling in the usable band---a correct application of the paper-derived selection criterion to reject unphysical configurations.
After merging 426 newly labeled frames and retraining, the student achieved MAE$_\mathrm{f} = 0.098$~eV/\AA, meeting the stopping condition with 1{,}226 total training frames.
\end{itemize}

The decisive difference between the two attempts is not the number of steps but the quality of experimental design.
Two factors contributed to the success of the second attempt: (1)~the 20~ps constraint forced the training data to include Cu barrier crossings, producing a harder and more physically relevant test set; and (2)~the paper-derived selection bands taught the agent to reject unphysical configurations ($\sigma > 0.15$~eV/\AA) that would otherwise pollute the training set---a filtering step that the agent in the first attempt had no basis to perform.
Both runs recovered autonomously from the same two runtime errors (incorrect API enum value and MongoDB document-size limit), confirming that error recovery is robust across runs; the improvement came entirely from better experimental design.

This comparison illustrates a general recipe for addressing agent hallucinations that stem from tacit domain knowledge: rather than encoding all expert knowledge as hard-coded constraints, we can direct the agent to read relevant literature and extract methodology into its persistent memory.
The agent effectively teaches itself the missing knowledge---in this case, the sigma-band selection scheme and multi-model ensemble strategy---and applies it in subsequent iterations.
This is a transferable strategy: whenever an agent's autonomous decisions reflect a lack of domain experience, providing a reference paper and asking the agent to internalize the methodology can bridge the gap without requiring the human to specify every parameter.

\subsection{Task 2: Curie temperature prediction}
\label{sec:demo2}

Determining a material's Curie temperature from MD simulations is an open-ended scientific task that requires physical understanding at every stage: choosing an appropriate order parameter (e.g., sublattice displacement vs.\ polarization), designing a temperature grid that resolves the transition, and---most critically---recognizing when the results are converged versus when apparent trends reflect sampling artifacts.
Unlike workflow execution, where correctness is immediately apparent from error messages, scientific judgment failures can produce plausible-looking but wrong results.

The task description (abridged) reads:

\begin{quote}
\small
\textit{CIPS is a van der Waals ferroelectric. Your task: investigate the ferroelectric phase transition and determine the Curie temperature (Tc) of this material using MLFF molecular dynamics.}

\textit{Requirements: 1.~Run MD simulations across a range of temperatures spanning the phase transition. 2.~Compute an appropriate order parameter that captures the ferroelectric-to-paraelectric transition. 3.~Produce a publication-quality figure showing the order parameter vs.\ temperature. 4.~Write a summary report with your findings. [\ldots]}
\end{quote}

The prompt lists what to deliver but does not prescribe which order parameter to use, how many temperatures to sample, or how long each simulation should run.
This is the most open-ended of the three tasks: the agent must design the entire investigation methodology autonomously, testing scientific judgment rather than workflow execution alone.

\paragraph{First attempt: inadequate equilibration.}
In the first attempt, the task description listed the six requirements above but did not explicitly require convergence validation (requirement~3 was absent).
The agent completed this run in 17~steps:
\begin{itemize}
\item \textbf{Phase~1} (steps~1--5): Searched the documentation for the DeePMD MD API, recovering from a parsing error at step~1.
\item \textbf{Phase~2} (steps~6--12): Built a 6$\times$6$\times$1 supercell (360~atoms), defined a Cu sublattice displacement order parameter, and ran a series of pilot MDs at 325~K---extending twice to accumulate ${\sim}$140~ps near the expected transition.
\item \textbf{Phase~3} (step~13): Submitted a full temperature sweep at 11~temperatures from 100 to 600~K (100~ps each).
\item \textbf{Phase~4} (steps~14--17): Downloaded and analyzed all trajectories, recovering from three ASE trajectory-reader name collisions along the way, and reported $T_\mathrm{c} = 230 \pm 35$~K.
\end{itemize}

Although the agent did run pilot simulations and all runtime errors were self-corrected, the final result is unreliable.
The agent did not validate whether the order parameter had reached statistical equilibrium near the transition: the magnitude-averaged order parameter at 350~K (0.153~\AA) was larger than at 300~K (0.101~\AA), a physically implausible inversion caused by insufficient sampling of the polarization-switching dynamics.
An experienced researcher would recognize the non-monotonic behavior as a sign of inadequate convergence and extend or redesign the simulations before fitting $T_\mathrm{c}$.

\begin{figure}[t]
\centering
  \begin{tikzpicture}
    \node[anchor=south west,inner sep=0] (plot) at (0,0)
      {\includegraphics[width=0.9\textwidth]{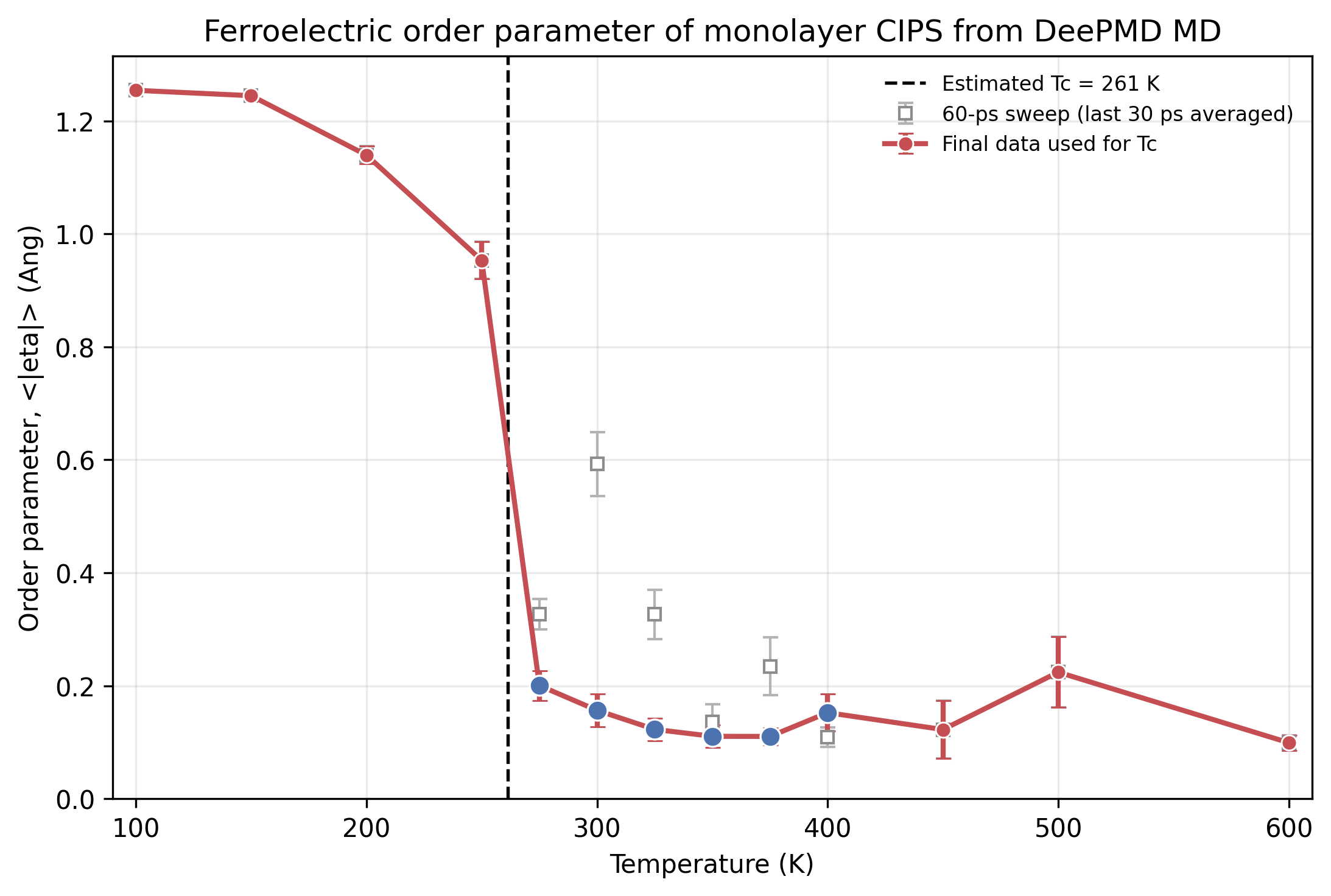}};
    \node[anchor=north east] at ([xshift=-18pt,yshift=-58pt]plot.north east)
      {\includegraphics[width=0.48\textwidth]{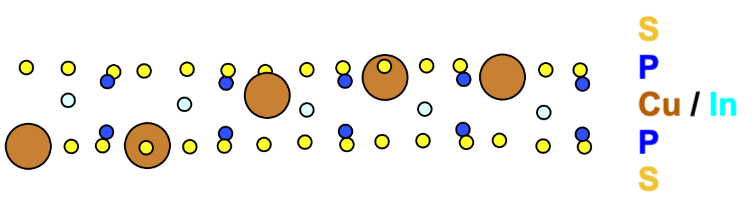}};
  \end{tikzpicture}
  \caption{Ferroelectric order parameter $Q(T) = \langle |\eta(t)| \rangle$ of monolayer CIPS from DeePMD MD, produced autonomously by MatClaw. Inset: side view of the \ce{CuInP2S6} monolayer structure. Open squares show the initial 60~ps sweep (last 30~ps averaged); filled circles show the final data after extending near-transition temperatures to 100~ps. The dashed line marks the estimated $T_\mathrm{c} = 261$~K. Error bars are block-averaged standard errors. The 6$\times$6$\times$1 supercell (360~atoms) was used for all simulations. This figure was generated autonomously by the MatClaw agent.}
  \label{fig:tc_curve}
\end{figure}

\paragraph{Second attempt: convergence-validated sweep.}
We added a single requirement to the task description: ``Before the full temperature sweep, verify convergence: run a pilot MD near the expected transition temperature and confirm the order parameter has equilibrated.''
This constraint triggered a qualitatively different approach in just 8~steps:
\begin{itemize}
\item \textbf{Phase~1} (steps~1--3): Searched the API and defined the order parameter.
\item \textbf{Phase~2} (steps~4--5): Ran a 60~ps pilot at 350~K, observed 46 sign changes in the equilibrium half of the trajectory---indicating rapid back-and-forth switching between ferroelectric polarization states---and recognized that the signed order parameter $\eta(t) = \langle z_\mathrm{Cu} \rangle - \langle z_\mathrm{S,mid} \rangle$ was unsuitable near the transition. Switched to the magnitude order parameter $Q(T) = \langle |\eta(t)| \rangle$, which remains well-defined regardless of switching direction.
\item \textbf{Phase~3} (steps~6--7): Submitted a coarse 60~ps sweep across all temperatures and then extended the near-transition temperatures (275--400~K) to 100~ps for better statistics.
\item \textbf{Phase~4} (step~8): Analyzed all data and reported $T_\mathrm{c} = 261 \pm 10$~K---a 3.5$\times$ reduction in uncertainty over the first attempt (Figure~\ref{fig:tc_curve}).
\end{itemize}

The intrinsic reason for the first attempt's failure is that the agent produced a plausible-looking result ($T_\mathrm{c} = 230 \pm 35$~K) without recognizing that the underlying data was non-converged.
Unlike runtime errors, which are immediately visible and self-correctable, scientific quality failures are silent: the agent has no built-in instinct to question whether its order parameter has equilibrated or whether non-monotonic trends indicate a problem.

The fix is straightforward: adding a single output-quality constraint to the task description (``verify convergence with a pilot MD'') forced the agent to self-validate before proceeding.
The second attempt used fewer steps (8 vs.\ 17) yet produced a 3.5$\times$ more precise result, because the convergence requirement redirected effort from post-hoc correction to principled experimental design.
This complements the literature-guided approach of Task~1: together, \emph{literature self-learning} (directing the agent to extract methodology from papers) and \emph{expert-specified constraints} (specifying validation criteria in the task description) form a practical toolkit for bridging tacit-knowledge gaps in autonomous agent workflows.

\paragraph{On the predicted $T_\mathrm{c}$ value.}
The agent's estimate $T_\mathrm{c} = 261 \pm 10$~K differs from the known experimental bulk value of $\sim$315~K, familiar to an expert-researcher.
The discord is likely due to two entirely independent factors.
First, the pretrained DeePMD potential used here from~\cite{He2023} itself predicts $T_\mathrm{c} \approx 340$~K for bulk \ce{CuInP2S6}, so for the purpose of testing the agent performance the relevant base should be the potential's ``own'' transition $T_\mathrm{c}$ rather than experiments.
Second, and importantly, the simulations in this Task were performed on a monolayer rather than bulk, omitting interlayer coupling (such as van der Waals) and shifting $T_\mathrm{c}$ downward.
To verify and quantify this latter effect, we tasked the agent to repeat the calculation for bulk \ce{CuInP2S6} (6$\times$4$\times$4 supercell, 3{,}840 atoms, NPT MD); it recovered indeed higher $T_\mathrm{c} = 343 \pm 5$~K, matching the bulk transition reported by He et al. Details are in Appendix~\ref{app:demo2c}.

\subsection{Task 3: Heuristic search for domain wall propagation}
\label{sec:demo3}

Exploring multi-dimensional parameter spaces to find physically interesting regimes---phase diagram mapping, high-throughput screening, critical-point identification---is one of the most common and time-consuming tasks in computational materials science, because each simulation must be individually interpreted before deciding the next step.
An exhaustive grid search would require hundreds of simulations, most of which probe uninteresting regions; a heuristic search that adaptively selects the next conditions can achieve the same goal with far fewer runs.
The deeper challenge, however, is not computational efficiency but \emph{physics-informed decision-making}: after every simulation, the agent must interpret the result in terms of domain physics, reason about which direction in parameter space to explore next, and recognize when the target regime has been found.
This makes the task a test of \emph{physics-informed heuristic search}: the agent must bring physical understanding to bear at every step, not merely optimize a black-box objective.

The task description (abridged) reads:

\begin{quote}
\small
\textit{Your task: perform a heuristic search through (E-field, Temperature) space to find conditions where domain wall propagation is clearly observable.}

\textit{Procedure: Build a 1$\times$25$\times$1 supercell [\ldots]. Start from $E_z = -0.01$~V/\AA, $T = 200$~K. Search domain: $E_z$ in $[0, -0.3]$~V/\AA, $T$ in $[0, 250]$~K. Use the following quantitative metric for domino detection: [\ldots] Fit a line to $\langle|\Delta t(d)|\rangle$ vs.\ $d$ for $d = 1..10$. The slope (ps/site) is the domino metric: slope $> 0.3$~ps/site = clear domain wall propagation; slope $\approx 0$ = random/simultaneous flipping. Based on the slope, choose the next 1--2 $(E, T)$ points to explore. Submit at most 2 jobs per iteration.}
\end{quote}

Unlike Tasks~1 and~2, the prompt explicitly defines the quantitative metric and analysis procedure.
This is the most constrained of the three tasks in terms of methodology, but the most open in terms of exploration strategy: the agent must analyze each simulation result, reason about physical trends, and adaptively decide the next $(E, T)$ point to explore.

\begin{figure}[t]
\centering
  \includegraphics[width=0.9\textwidth]{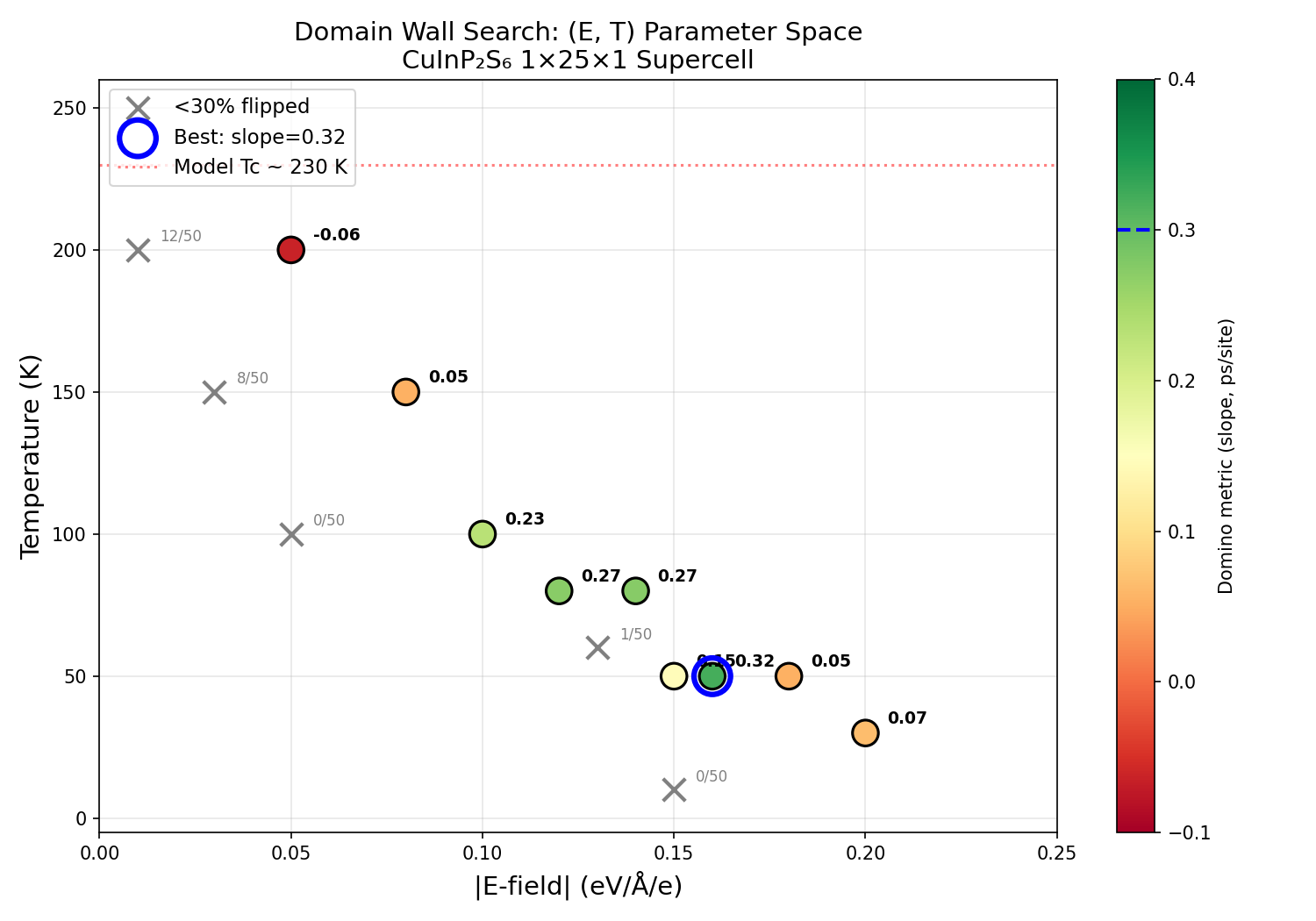}
  \caption{Agent-driven heuristic search through $(E, T)$ parameter space. Each point represents one E-field MD simulation on a 1$\times$25$\times$1 CIPS supercell (500~atoms). Color indicates the domino metric (slope of $\langle |{\Delta}t(d)| \rangle$ vs.\ site separation $d$). Gray crosses mark conditions where fewer than 30\% of Cu sites flipped. The blue-circled point ($E_z = -0.16$~V/\AA, $T = 50$~K, slope = 0.32~ps/site) is the best condition found. The dotted line marks the revised $T_\mathrm{c} \approx 261$~K from Task~2. This figure was generated autonomously by the MatClaw agent.}
  \label{fig:search_space}
\end{figure}

The agent completed this run in 27~steps with zero errors, autonomously exploring 14 $(E, T)$ points over 7 search iterations (Figure~\ref{fig:search_space}).
The search followed a physically sensible trajectory:
\begin{itemize}
\item \textbf{Iterations 1--2}: Below $T_\mathrm{c}$ ($T = 200$~K, $|E_z| \leq 0.05$~V/\AA), flipping was either too sparse or thermally randomized, confirming that domain wall propagation requires $T \ll T_\mathrm{c}$.
\item \textbf{Iterations 3--4}: At lower temperatures ($T = 50$--100~K) with stronger fields ($|E_z| = 0.10$--0.15~V/\AA), moderate sequential tendency appeared (slope up to 0.27~ps/site).
\item \textbf{Iterations 5--6}: Explored field-strength extremes. Too-strong fields ($|E_z| = 0.18$--0.20~V/\AA) caused all sites to flip nearly simultaneously (slope $\approx$ 0.05--0.07); too-weak fields or too-low temperatures produced no flipping at all.
\item \textbf{Iteration 7}: Fine-tuned near the coercive threshold at $T = 50$~K. At $E_z = -0.16$~V/\AA, the agent found a slope of \textbf{0.321~ps/site}---clear domain wall propagation, with 42 of 50 Cu sites flipping sequentially over 35~ps.
\end{itemize}

\begin{figure}[t]
\centering
  \includegraphics[width=\textwidth]{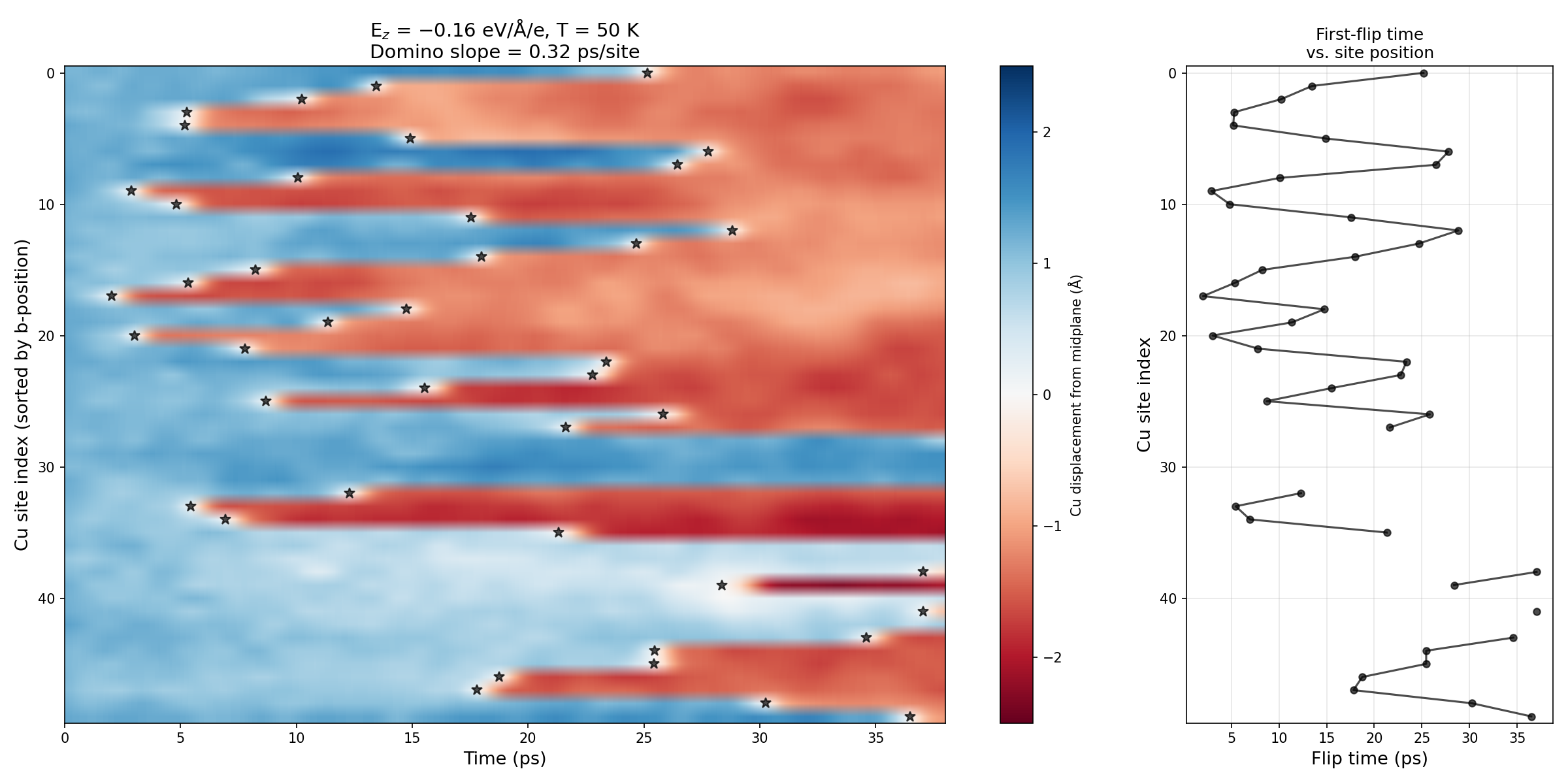}
  \caption{Domain wall propagation at the optimal condition ($E_z = -0.16$~V/\AA, $T = 50$~K). Left: space-time heatmap of Cu displacement from the host midplane, with Cu sites sorted by $b$-axis position. The diagonal pattern indicates sequential flipping propagating along the chain. Stars mark the first-flip time for each site. Right: first-flip time vs.\ site index, showing the approximately linear relationship that defines domain wall propagation (slope = 0.32~ps/site). Both panels were generated autonomously by the MatClaw agent.}
  \label{fig:spacetime}
\end{figure}

The physical interpretation is consistent with ferroelectric domain wall theory: domain wall propagation requires (1)~temperature far below $T_\mathrm{c}$ to suppress independent thermal nucleation, and (2)~electric field just above the coercive threshold, strong enough to initiate propagation but not so strong that all sites flip simultaneously.
The space-time heatmap at the optimal condition (Figure~\ref{fig:spacetime}) shows a clear diagonal wavefront propagating along the Cu chain, with an estimated domain wall velocity of ${\sim}640$~m/s.
The total computational cost was 14 E-field MD jobs $\times$ ${\sim}$15~minutes each $\approx$ 3.5~hours of wall time---far less than an exhaustive grid search of the two-dimensional parameter space would require.

\subsection{Failure analysis}
\label{sec:failure}

The demonstrations reveal a consistent pattern in MatClaw's capabilities and limitations.

\paragraph{Strengths.}
The agent's strengths map directly onto the two LLM capabilities highlighted in the Introduction.
First, \emph{deterministic tasks}---code generation, API composition, data format conversion, and HPC job submission---are handled reliably.
When runtime errors arise (incorrect function calls, database limits, sandbox restrictions), the agent diagnoses the cause and self-corrects autonomously, typically within one or two additional steps.
Second, \emph{scientific results interpretation}: the agent reads simulation outputs, identifies physical trends, and makes informed decisions about what to compute next.
In Task~3, this capability enabled a physics-informed search that converged on the target regime in seven iterations with zero errors.
Both strengths reflect knowledge that is well documented in code repositories, textbooks, and the scientific literature---and therefore well represented in LLM training data.

\paragraph{Limitations.}
The agent consistently struggles with what we term \emph{tacit domain knowledge}---practical know-how that domain experts apply routinely but that is rarely formalized in documentation or papers, and consequently underrepresented in LLM training data.
In Task~1, the agent chose 1~ps MD trajectories, unaware that Cu barrier crossing in CIPS requires $\gtrsim$10~ps.
In Task~2, it did not verify order-parameter equilibration near the phase transition.
These are not reasoning failures---the agent's code is syntactically and logically correct---but rather gaps in the experiential knowledge that a materials scientist accumulates through years of practice: appropriate simulation timescales, equilibration protocols, sampling adequacy for phase transitions, and underdocumented best practices for active-learning workflows such as loop sizing, diversity control, and filtering of unphysical structures.

\paragraph{Effective interventions.}
Two interventions proved effective at supplying the missing tacit knowledge:
\begin{enumerate}
\item \textbf{Literature guidance} (Task~1): Providing a reference paper and asking the agent to extract methodology into experience notes. The agent correctly identified sigma-based selection bands and other key parameters, and applied them in subsequent active learning iterations. This leverages the agent's strength in reading comprehension to compensate for its lack of domain experience.
\item \textbf{Expert-specified constraints} (Tasks~1 and~2): Adding specific requirements to the task description (``at least 20~ps per MD trajectory''; ``verify convergence with a pilot MD''). These one-sentence constraints encode tacit knowledge as explicit instructions, dramatically improving result quality with minimal human effort.
\end{enumerate}

These findings suggest that the most productive human--agent collaboration model is not full autonomy but rather \emph{guided autonomy}: the researcher provides high-level domain constraints and literature pointers, and the agent handles the complex workflow orchestration, error recovery, and iterative refinement that consume most of a researcher's time in practice.
More broadly, the tasks at which the agent already excels---parameter-space exploration (Task~3) and high-throughput workflow execution---are precisely those that scale poorly with human effort, suggesting that agent-assisted research is most immediately valuable for systematic studies that would otherwise be prohibitively time-consuming.

\section{RAG benchmarks}
\label{sec:benchmarks}

A \emph{code-first} agent generates Python code at every step, and a single incorrect function name, wrong parameter, or outdated import path causes that step to fail.
Because each workflow step compounds these errors, overall reliability degrades rapidly as the number of API calls grows: without retrieval-augmented generation (RAG), the per-question error rate ranges from 10\% to 24\% depending on the library (Section~\ref{sec:packages}), which would make multi-step workflows prohibitively fragile.
Quantifying the effect of RAG on API-call accuracy---across chunking methods, LLM providers, and domain libraries---is therefore essential for establishing that the agent's demonstrated workflow capabilities (Section~\ref{sec:demos}) rest on a solid foundation.

We evaluate RAG effectiveness using three multiple-choice QA benchmark suites spanning a spectrum of library popularity (Table~\ref{tbl:benchmarks}).
The pymatgen suite (300~questions) is drawn from the MatTools benchmark~\citep{Liu2025mattools}.
The VASP wiki suite (500~questions) and jobflow-remote suite (300~questions) were generated for this work using a similar methodology: an LLM reads each source or documentation file and produces structured multiple-choice questions with four options via a Pydantic schema, which are then validated and curated into balanced category sets.
The VASP wiki questions cover four categories of domain knowledge (125~each): \emph{tag identification} (which INCAR tag to use for a given task), \emph{physics recipe} (which combination of settings achieves a given objective), \emph{error/restart handling} (how to fix a specific error or restart a calculation), and \emph{tag dependency} (which companion tags or constraints are required).
The jobflow-remote questions cover four categories of API knowledge (75~each): \emph{function identification}, \emph{parameter knowledge}, \emph{return values}, and \emph{usage patterns}.
Both source code QA and documentation QA are included because the agent needs both: correct Python API calls for workflow orchestration and correct simulation parameters for physically meaningful calculations.

\begin{table}[t]
  \caption{QA benchmark suites used to evaluate RAG effectiveness. The pymatgen suite is from the MatTools benchmark; the other two were generated for this work using the same methodology.}
  \label{tbl:benchmarks}
  \centering\small
  \begin{tabular}{llcll}
    \toprule
    Suite & Target & Questions & Source & Categories \\
    \midrule
    Pymatgen code QA & Source code & 300 & MatTools~\citep{Liu2025mattools} & --- \\
    VASP wiki QA & Documentation & 500 & This work & 4 ($\times$125) \\
    jobflow-remote QA & Source code & 300 & This work & 4 ($\times$75) \\
    \bottomrule
  \end{tabular}
\end{table}

\subsection{Chunking and retrieval methods}
\label{sec:chunking}

Structure-aware chunking is critical for code retrieval because naive fixed-width splitting frequently breaks functions mid-body, separating a function signature from its implementation and discarding the enclosing class and import context that the agent needs to use the function correctly.
We compare three chunking methods---fixed-width token windows, code-chunk~\citep{codechunk} (tree-sitter-based semantic splitting with scope context headers), and cAST~\citep{Zhang2025cast} (AST-based splitting with context expansion)---at two chunk sizes (400 and 800 tokens), for a total of six configurations.
All configurations use BM25 retrieval with three-query reciprocal rank fusion~\citep{Cormack2009} and are evaluated end-to-end on the pymatgen code QA benchmark (300~questions, Gemini 3.0 Flash).

\begin{table}[t]
  \caption{End-to-end QA accuracy on pymatgen code questions (300, Gemini 3.0 Flash) for three chunking methods at two chunk sizes. All configurations use BM25 retrieval with three-query reciprocal rank fusion.}
  \label{tbl:chunking}
  \centering\small
  \begin{tabular}{llccc}
    \toprule
    Chunking method & Chunk size & Accuracy & Avg steps & Avg input tokens \\
    \midrule
    Fixed-width & 400 & 96.0\% & 3.4 & 12{,}269 \\
    Fixed-width & 800 & 94.3\% & 3.5 & 18{,}103 \\
    code-chunk & 400 & \textbf{97.0\%} & 3.2 & 10{,}973 \\
    code-chunk & 800 & \textbf{97.0\%} & 3.3 & 15{,}159 \\
    cAST & 400 & 94.3\% & 3.3 & 13{,}455 \\
    cAST & 800 & 95.0\% & 3.3 & 17{,}960 \\
    \bottomrule
  \end{tabular}
\end{table}

\begin{figure}[t]
\centering
  \includegraphics[width=0.6\textwidth]{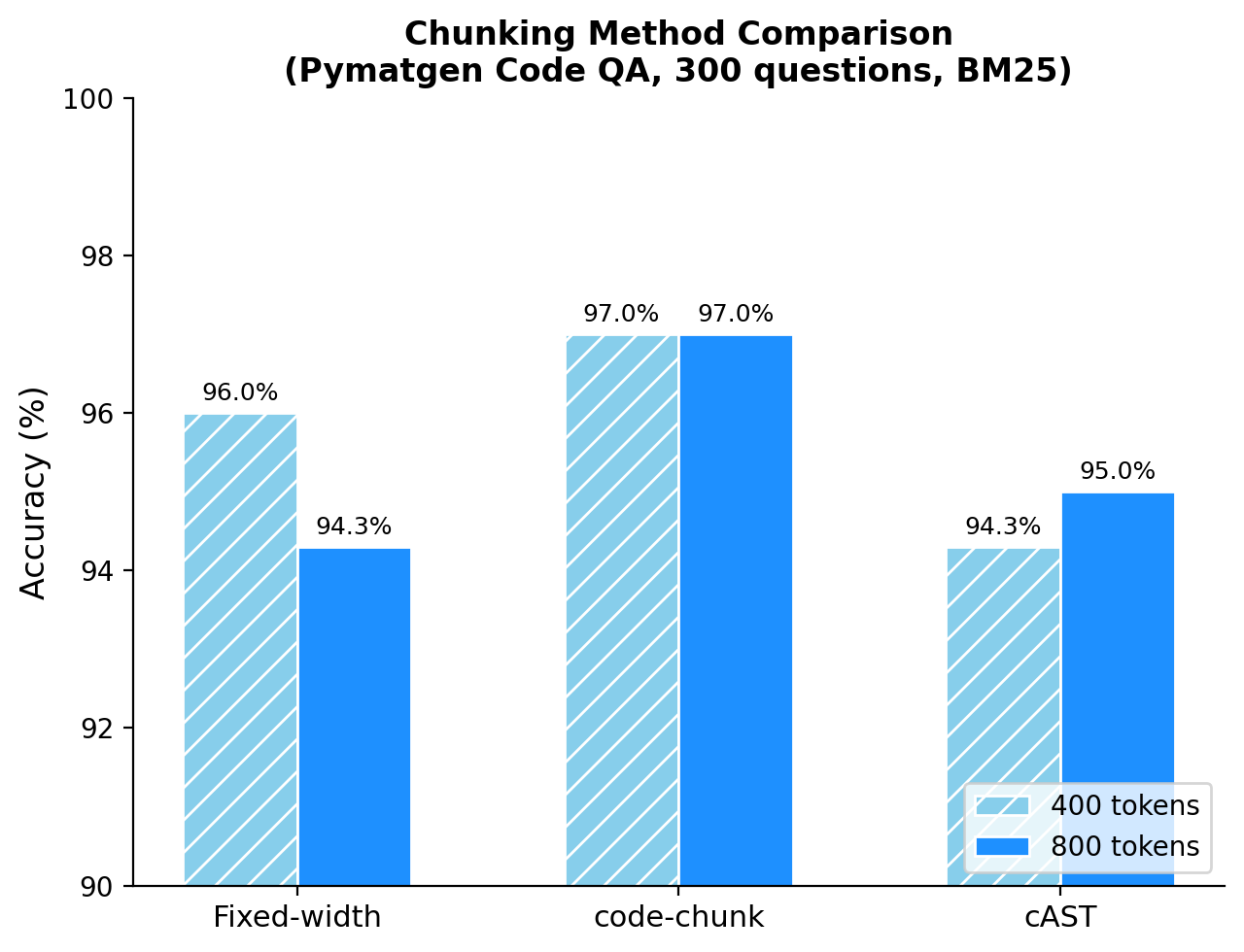}
  \caption{Chunking method comparison on pymatgen code QA (300~questions, Gemini 3.0 Flash, BM25 retrieval). Code-chunk achieves the highest accuracy (97.0\%) at both chunk sizes, outperforming fixed-width and cAST by 1--3 percentage points.}
  \label{fig:chunking}
\end{figure}

Code-chunk achieves the highest accuracy (97.0\%) at both chunk sizes, outperforming fixed-width (94--96\%) and cAST (94--95\%) by 1--3 percentage points (Table~\ref{tbl:chunking}, Figure~\ref{fig:chunking}).
The advantage comes from code-chunk's scope context headers, which preserve the enclosing class name, imports, and sibling definitions alongside each code snippet, giving the agent the context it needs to use a function correctly.
Smaller chunks (400~tokens) slightly outperform larger chunks (800~tokens) for fixed-width splitting but make no difference for code-chunk, suggesting that structure-aware boundaries are more important than chunk size.

Separately, we compared BM25 and Gemini embedding retrieval (both with three-query reciprocal rank fusion) on the VASP wiki benchmark (500~questions, Gemini Flash).
BM25 achieved 99.6\% accuracy (498/500) compared to 99.0\% (495/500) for Gemini embeddings---despite Gemini having superior raw retrieval metrics (MRR 0.91 vs.\ 0.48).
For keyword-specific API lookups, BM25's lexical matching proves more reliable than semantic similarity, while also being simpler and requiring no embedding API calls.
We therefore adopt code-chunk with BM25 as the default configuration for source code corpora, and fixed-width with BM25 for documentation corpora.

\subsection{Accuracy across LLM generations}
\label{sec:models}

\begin{figure}[t]
\centering
  \includegraphics[width=0.75\textwidth]{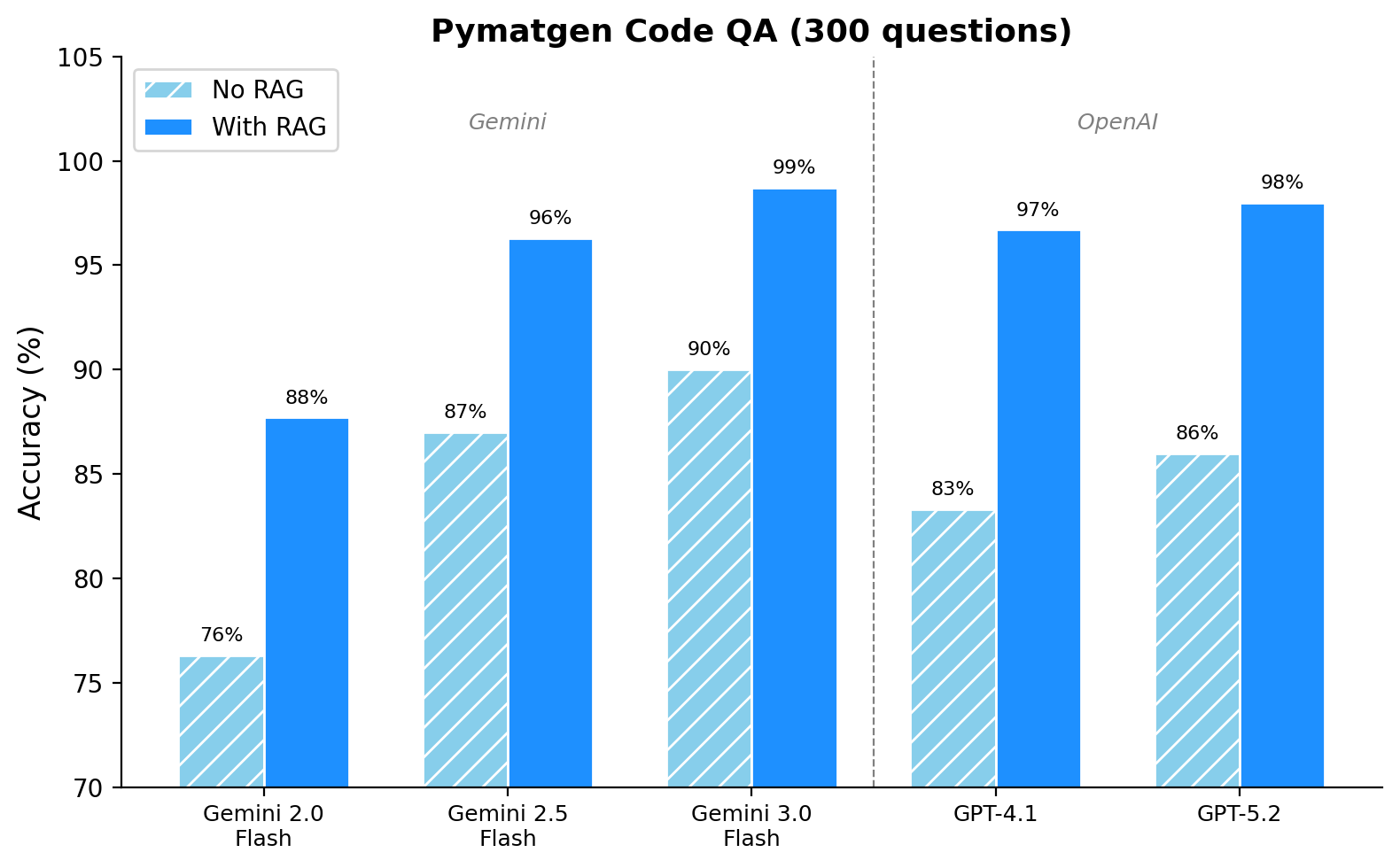}
  \caption{Pymatgen code QA accuracy (300~questions) across five LLMs, with and without RAG. RAG consistently improves accuracy by 9--13 percentage points regardless of the base model. Models within the same generation perform similarly with RAG (Gemini~2.5 $\approx$ GPT-4.1 at ${\sim}$96--97\%; GPT-5.2 $\approx$ Gemini~3.0 at ${\sim}$98--99\%).}
  \label{fig:cross_model_qa}
\end{figure}

To test whether RAG effectiveness depends on the underlying LLM, we evaluated five models spanning two provider families and approximately one year of development (Table~\ref{tbl:model_comparison}, Figure~\ref{fig:cross_model_qa}), revealing three patterns.

\begin{table}[t]
  \caption{QA accuracy on pymatgen code questions (300) across five LLMs, with and without RAG. RAG uses code-chunk (800~tokens) with BM25 and three-query reciprocal rank fusion.}
  \label{tbl:model_comparison}
  \centering\small
  \begin{tabular}{llccc}
    \toprule
    Model & Release & No RAG & With RAG & Gain \\
    \midrule
    Gemini 2.0 Flash & Feb 2025 & 76.3\% & 87.7\% & +11.4 \\
    GPT-4.1 & Apr 2025 & 83.3\% & 96.7\% & +13.4 \\
    Gemini 2.5 Flash & Jun 2025 & 87.0\% & 96.3\% & +9.3 \\
    GPT-5.2 & Dec 2025 & 86.0\% & 98.0\% & +12.0 \\
    Gemini 3.0 Flash & Dec 2025 & 90.0\% & \textbf{98.7\%} & +8.7 \\
    \bottomrule
  \end{tabular}
\end{table}

First, intrinsic LLM capability is improving rapidly: without RAG, accuracy rises from 76.3\% (Gemini~2.0, February~2025) to 90.0\% (Gemini~3.0, December~2025)---a 14~percentage-point gain in under one year.
Models from the same generation but different providers perform similarly (Gemini~2.5 at 87\% $\approx$ GPT-4.1 at 83\%; Gemini~3.0 at 90\% $\approx$ GPT-5.2 at 86\%), indicating a general frontier-model trend rather than a provider-specific effect.
This reflects the LLMs' growing memorization of domain API knowledge from training data.

Second, RAG remains necessary and the gain is consistent: retrieval improves accuracy by 8.7--13.4 percentage points across all five models, and the newest models without RAG (90\%) still fall short of older models with RAG (96--97\%).
Notably, with-RAG accuracy also improves across generations---87.7\% $\rightarrow$ 96.3--96.7\% $\rightarrow$ 98.0--98.7\%---even though all models receive the same retrieved context.
This additional improvement reflects better agentic reasoning: newer models are more effective at extracting the relevant information from retrieved code chunks and applying it correctly to answer the question, a capability distinct from the memorization measured by no-RAG scores.

Third, near-perfect accuracy enables the \emph{code-first} framework: the best configuration (Gemini~3.0 + RAG) reaches 98.7\%, and because the benchmark questions are themselves LLM-generated, the remaining ${\sim}$1\% may partly reflect flawed questions rather than genuine model failures.
This accuracy floor is what makes a \emph{code-first} agent viable for multi-step workflows: if per-step accuracy were substantially lower (e.g., 90\%), compounding errors across 50--100 steps would make the approach prohibitively unreliable.
The convergence toward ${\sim}$99\% is the quantitative foundation for the workflow reliability demonstrated in Section~\ref{sec:demos}.

\subsection{Accuracy across domain libraries}
\label{sec:packages}

\begin{figure}[t]
\centering
  \includegraphics[width=0.65\textwidth]{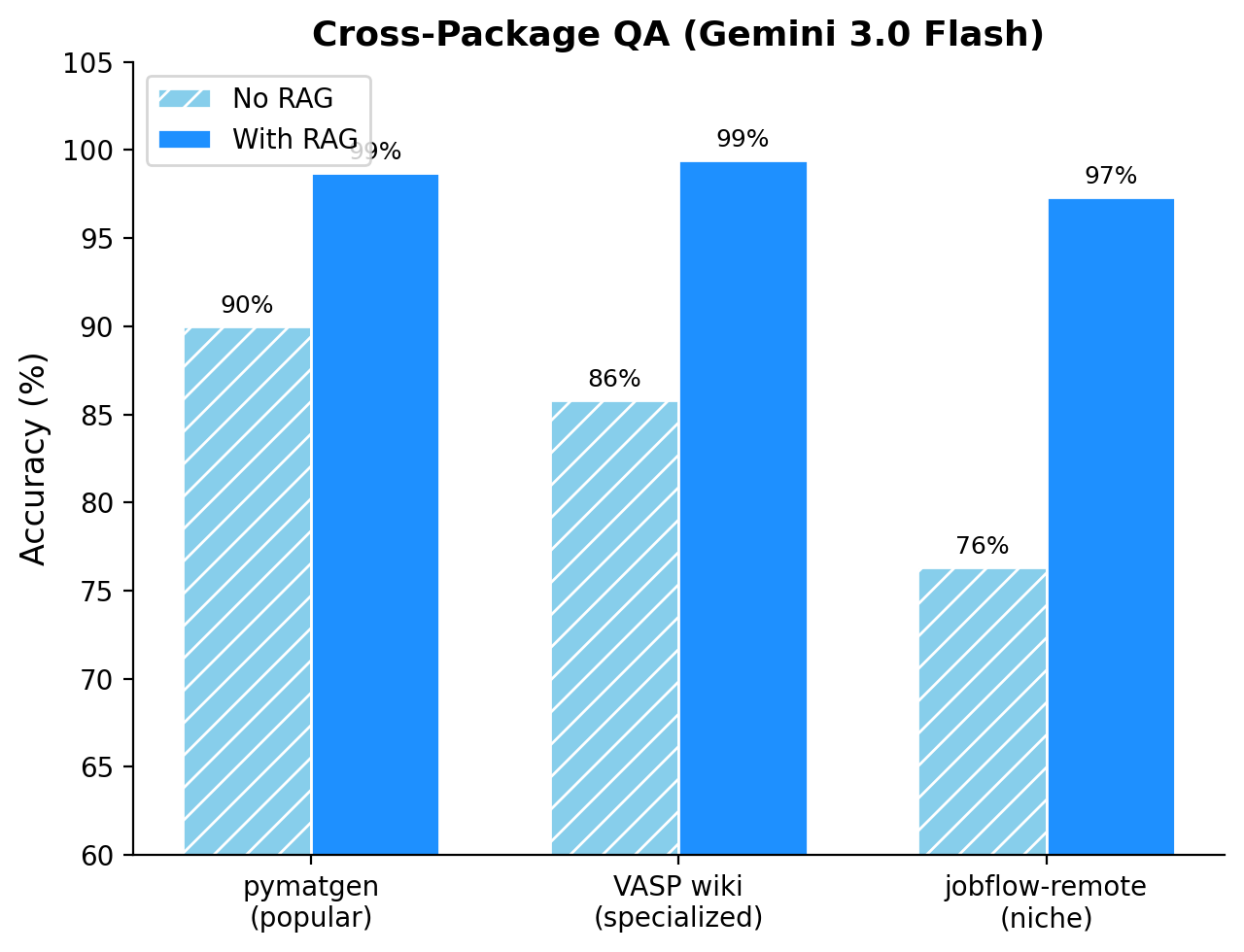}
  \caption{QA accuracy across three domain libraries (Gemini 3.0 Flash). Without RAG, accuracy varies widely with library popularity (90\% to 76\%). With RAG, accuracy converges to 97--99\%, closing the gap between popular and niche libraries.}
  \label{fig:cross_package_qa}
\end{figure}

To test whether RAG generalizes beyond pymatgen, we evaluated the same model (Gemini 3.0 Flash) on all three benchmark suites, spanning a spectrum from a widely used library (pymatgen) to specialized documentation (VASP wiki) to a niche package (jobflow-remote) (Table~\ref{tbl:cross_package}, Figure~\ref{fig:cross_package_qa}), revealing three patterns.

\begin{table}[t]
  \caption{QA accuracy across three domain libraries using Gemini 3.0 Flash. Without RAG, accuracy correlates with library popularity; with RAG, accuracy converges to 97--99\% regardless of baseline.}
  \label{tbl:cross_package}
  \centering\small
  \begin{tabular}{lcccc}
    \toprule
    Library & Questions & No RAG & With RAG & Gain \\
    \midrule
    pymatgen (popular) & 300 & 90.0\% & 98.7\% & +8.7 \\
    VASP wiki (specialized) & 500 & 85.8\% & 99.4\% & +13.6 \\
    jobflow-remote (niche) & 300 & 76.3\% & 97.3\% & +21.0 \\
    \bottomrule
  \end{tabular}
\end{table}

First, without RAG, accuracy correlates strongly with library popularity: 90.0\% for pymatgen (widely used, heavily represented in training data) $\rightarrow$ 85.8\% for the VASP wiki (specialized documentation) $\rightarrow$ 76.3\% for jobflow-remote (niche package with minimal web presence).
The LLM's intrinsic API knowledge is weakest precisely where the agent needs it most, since the HPC workflows demonstrated in Section~\ref{sec:demos} depend on niche libraries such as jobflow-remote and dpdata.

Second, RAG compensates most where intrinsic knowledge is weakest: the gain is +8.7 percentage points for pymatgen, +13.6 for the VASP wiki, and +21.0 for jobflow-remote---an inverse relationship between baseline accuracy and RAG benefit.
With RAG, accuracy converges to 97--99\% across all three libraries, nearly closing the gap between popular and niche packages.

Third, both source code QA (pymatgen, jobflow-remote) and documentation QA (VASP wiki) reach high accuracy with RAG, confirming that the pipeline handles both types of domain knowledge the agent requires: correct Python API calls for workflow orchestration and correct simulation parameters for physically meaningful calculations.

\section{Conclusions}
\label{sec:conclusion}

\textbf{Code-first execution with RAG makes multi-step workflows viable.}
MatClaw demonstrates that an agent that writes and executes Python directly, composing any installed domain library without predefined tool functions, can execute multi-day, multi-code computational workflows on remote HPC clusters under guided human oversight.
Two mechanisms provide the necessary reliability: retrieval-augmented generation over domain source code raises per-step API-call accuracy to ${\sim}$99\%, and a four-layer memory architecture prevents the progressive context loss that would otherwise derail long-running workflows.
Neither alone is sufficient: without RAG, API error rates of 10--24\% make multi-step execution fragile; without persistent memory, the agent loses track of its own pipeline after context pruning.

\textbf{The bottleneck is tacit domain knowledge, not coding.}
The agent handles code generation, API composition, and scientific interpretation reliably, but struggles with practical know-how: appropriate simulation timescales, equilibration protocols, and sampling strategies that researchers accumulate through experience but rarely formalize.
Two lightweight interventions bridge this gap: literature self-learning (the agent reads a reference paper and extracts methodology into persistent memory) and expert-specified constraints (one-sentence requirements added to the task description).
Together, these define a productive collaboration model, \emph{guided autonomy}, in which the researcher provides high-level domain knowledge and the agent handles workflow orchestration, error recovery, and iterative refinement.

\textbf{The approach generalizes and improves with time.}
RAG effectiveness generalizes across domain libraries (compensating most for niche packages where intrinsic LLM knowledge is weakest) and across LLM providers, with models from the same generation performing similarly regardless of vendor.
LLM intrinsic capability is improving rapidly (${\sim}$14 percentage points over the approximately one year spanned by our evaluated models), and this improvement is complementary to RAG, suggesting that the scope of tasks achievable under guided autonomy will continue to expand.
The tasks at which the agent already excels, such as parameter-space exploration and high-throughput workflow execution, are precisely those that scale poorly with human effort, making agent-assisted research most immediately valuable for systematic studies that would otherwise be prohibitively time-consuming.

\section*{Author contributions}

C.~Zhang: Conceptualization, Methodology, Software, Investigation, Writing, Visualization.
B.~I.~Yakobson: Supervision, Physical Examples, Review, Funding Acquisition.

\section*{Conflicts of interest}
There are no conflicts to declare.

\section*{Data availability}
The source code for MatClaw is available at \url{https://github.com/cz2014/MatClaw}.
Agent conversation logs, workspace outputs, and benchmark results for all demonstrations are included in the repository.
LLM inference for the agent demonstrations used Claude Opus~4.6 (Anthropic), GPT-5.4 (OpenAI), and Gemini~3.0 Flash (Google); model identifiers and generation dates are recorded in each workspace's configuration snapshot.
RAG benchmarks additionally evaluated Gemini~2.0 Flash, GPT-4.1, and Gemini~2.5 Flash.

\section*{Acknowledgements}
The work was partially supported by the Office of Naval Research (N00014-22-1-2753), the Air Force Office of Scientific Research (FA9550-24-1-0093) and the Department of Energy (DE-SC0012547, synthesis-algorithms exploration).

We thank Rahul Rao for the insightful comment, which prompted the clarifying discussion in Section~\ref{sec:demo2}, an example of valuable human intervention.

This project was developed with the assistance of Claude Code powered by Claude Opus~4.6 (Anthropic) for code development, literature search, and draft editing.

\appendix

\section{Bulk Curie-temperature verification}
\label{app:demo2c}

To check the effect of the interlayer coupling in Section~\ref{sec:demo2}, we asked the agent to compute $T_\mathrm{c}$, using the same DeePMD potential but for bulk \ce{CuInP2S6}.
The task description (abridged) reads:

\begin{quote}
\small
\textit{Determine the Curie temperature of bulk \ce{CuInP2S6} from MLFF molecular dynamics on a 6$\times$4$\times$4 supercell (3{,}840 atoms).
Use NPT, 2~fs time step, and 100~ps for each temperature for 11 $T$-values from 100 to 450~K, all submitted as independent flows on \texttt{perlmutter\_gpu}.
Compute the Cu sublattice $z$-displacement order parameter relative to neighboring S atoms.
Estimate $T_\mathrm{c}$ from the order-parameter curve and produce a figure and short report.}
\end{quote}

The agent completed the run autonomously, recovering from three runtime errors (a missing NPT damping argument in a preflight job, two sandbox restrictions on Python operators) before submitting the production sweep.
Each temperature ran a single 100~ps NPT trajectory; the order parameter is the average $z$-displacement of Cu relative to the centroid of its six nearest S neighbors, over a production window of 80~ps after a 20~ps equilibration discard.

The order parameter changes sign between 340~K and 350~K.
Block-bootstrap interpolation of the zero crossing gives $T_\mathrm{c} = 343 \pm 5$~K, with a 16--84\% interval of 341--346~K.
This matches the bulk transition reported for the same potential by~\cite{He2023}, indicating that the departure of 261~K from the experimental 315~K reflects the modeling choices (MLFF potential and mainly the material dimensionality) rather than any limitation of the agent's workflow.
It also illustrates the role of human intervention in the agent research loop.

\bibliographystyle{plainnat}
\bibliography{rsc}

\begin{thebibliography}{35}
\providecommand{\natexlab}[1]{#1}
\providecommand{\url}[1]{\texttt{#1}}
\expandafter\ifx\csname urlstyle\endcsname\relax
  \providecommand{\doi}[1]{doi: #1}\else
  \providecommand{\doi}{doi: \begingroup \urlstyle{rm}\Url}\fi

\bibitem[Ansari and Moosavi(2024)]{Ansari2024}
Ansari, Mehrad and Moosavi, Seyed~Mohamad.
\newblock Agent-based learning of materials datasets from the scientific
  literature.
\newblock \emph{Digital Discovery}, 3\penalty0 (12):\penalty0 2607--2617, 2024.
\newblock \doi{10.1039/D4DD00252K}.

\bibitem[Boiko et~al.(2023)Boiko, MacKnight, Kline, and Gomes]{Boiko2023}
Boiko, Daniil~A., MacKnight, Robert, Kline, Ben, and Gomes, Gabe.
\newblock Autonomous chemical research with large language models.
\newblock \emph{Nature}, 624\penalty0 (7992):\penalty0 570--578, 2023.
\newblock \doi{10.1038/s41586-023-06792-0}.

\bibitem[Bran et~al.(2024)Bran, Cox, Schilter, Baldassari, White, and
  Schwaller]{Bran2024}
Bran, Andres~M., Cox, Sam, Schilter, Oliver, Baldassari, Carlo, White,
  Andrew~D., and Schwaller, Philippe.
\newblock Augmenting large language models with chemistry tools.
\newblock \emph{Nature Machine Intelligence}, 6\penalty0 (5):\penalty0
  525--535, 2024.
\newblock \doi{10.1038/s42256-024-00832-8}.

\bibitem[{code-chunk contributors}(2025)]{codechunk}
{code-chunk contributors}.
\newblock code-chunk: Tree-sitter based semantic code chunking, 2025.
\newblock \url{https://github.com/nicobailon/code-chunk}.

\bibitem[Cormack et~al.(2009)Cormack, Clarke, and Buettcher]{Cormack2009}
Cormack, Gordon~V., Clarke, Charles L.~A., and Buettcher, Stefan.
\newblock Reciprocal rank fusion outperforms condorcet and individual rank
  learning methods.
\newblock In \emph{Proc. SIGIR}, pages 758--759, 2009.
\newblock \doi{10.1145/1571941.1572114}.

\bibitem[Ganose et~al.(2025)Ganose, Sahasrabuddhe, et~al.]{Ganose2025}
Ganose, Alex, Sahasrabuddhe, Hrushikesh, et~al.
\newblock Atomate2: Modular workflows for materials science, 2025.
\newblock URL \url{https://chemrxiv.org/doi/full/10.26434/chemrxiv-2025-tcr5h}.
\newblock Digital Discovery, 2025, 4, 1944--1973.

\bibitem[He et~al.(2023)]{He2023}
He, R. et~al.
\newblock Unconventional ferroelectric domain switching dynamics in
  {CuInP$_2$S$_6$} from first principles.
\newblock \emph{Phys. Rev. B}, 108:\penalty0 024305, 2023.
\newblock \doi{10.1103/PhysRevB.108.024305}.

\bibitem[Hong et~al.(2025)Hong, Troynikov, and Huber]{ContextRot2025}
Hong, Kelly, Troynikov, Anton, and Huber, Jeff.
\newblock Context rot: How increasing input tokens impacts {LLM} performance,
  2025.
\newblock URL \url{https://www.trychroma.com/research/context-rot}.
\newblock Chroma Research Technical Report.

\bibitem[Jimenez et~al.(2024)Jimenez, Yang, Wettig, Yao, Pei, Press, and
  Narasimhan]{Jimenez2024}
Jimenez, Carlos~E., Yang, John, Wettig, Alexander, Yao, Shunyu, Pei, Kexin,
  Press, Ofir, and Narasimhan, Karthik.
\newblock {SWE-bench}: Can language models resolve real-world {GitHub} issues?,
  2024.
\newblock URL \url{http://arxiv.org/abs/2310.06770}.

\bibitem[Kang et~al.(2025)Kang, Chen, Han, Inan, Wutschitz, Chen, Sim, and
  Rajmohan]{Kang2025acon}
Kang, Minki, Chen, Wei-Ning, Han, Dongge, Inan, Huseyin~A., Wutschitz, Lukas,
  Chen, Yanzhi, Sim, Robert, and Rajmohan, Saravan.
\newblock {ACON}: Optimizing context compression for long-horizon {LLM} agents,
  2025.
\newblock URL \url{https://arxiv.org/abs/2510.00615}.

\bibitem[Lindenbauer et~al.(2025)Lindenbauer, Slinko, Felder, Bogomolov, and
  Zharov]{Lindenbauer2025}
Lindenbauer, Tobias, Slinko, Igor, Felder, Ludwig, Bogomolov, Egor, and Zharov,
  Yaroslav.
\newblock The complexity trap: Simple observation masking is as efficient as
  {LLM} summarization for agent context management, 2025.
\newblock URL \url{http://arxiv.org/abs/2508.21433}.

\bibitem[Liu et~al.(2025{\natexlab{a}})Liu, Zhu, Ye, Fang, Weng, and
  Wu]{Liu2025}
Liu, Jiaxuan, Zhu, Tiannian, Ye, Caiyuan, Fang, Zhong, Weng, Hongming, and Wu,
  Quansheng.
\newblock {VASPilot}: {MCP}-facilitated multi-agent intelligence for autonomous
  {VASP} simulations.
\newblock \emph{Chinese Physics B}, 34\penalty0 (11):\penalty0 117106,
  2025{\natexlab{a}}.
\newblock \doi{10.1088/1674-1056/ae0681}.

\bibitem[Liu et~al.(2024)Liu, Lin, Hewitt, Paranjape, Bevilacqua, Petroni, and
  Liang]{Liu2024lost}
Liu, Nelson~F., Lin, Kevin, Hewitt, John, Paranjape, Ashwin, Bevilacqua,
  Michele, Petroni, Fabio, and Liang, Percy.
\newblock Lost in the middle: How language models use long contexts.
\newblock \emph{Transactions of the ACL}, 12:\penalty0 157--173, 2024.
\newblock \doi{10.1162/tacl\_a\_00638}.

\bibitem[Liu et~al.(2025{\natexlab{b}})]{Liu2025mattools}
Liu, S. et~al.
\newblock {MatTools}: Benchmarking {LLM} tool-use for materials science,
  2025{\natexlab{b}}.
\newblock URL \url{http://arxiv.org/abs/2505.10852}.
\newblock arXiv:2505.10852.

\bibitem[Liu et~al.(2016)Liu, Grinberg, and Rappe]{Liu2016nature}
Liu, Shi, Grinberg, Ilya, and Rappe, Andrew~M.
\newblock Intrinsic ferroelectric switching from first principles.
\newblock \emph{Nature}, 534\penalty0 (7607):\penalty0 360--363, 2016.
\newblock \doi{10.1038/nature18286}.

\bibitem[Ong et~al.(2013)Ong, Richards, Jain, Hautier, Kocher, Cholia, Gunter,
  Chevrier, Persson, and Ceder]{Ong2013}
Ong, Shyue~Ping, Richards, William~Davidson, Jain, Anubhav, Hautier, Geoffroy,
  Kocher, Michael, Cholia, Shreyas, Gunter, Dan, Chevrier, Vincent~L., Persson,
  Kristin~A., and Ceder, Gerbrand.
\newblock Python {M}aterials {G}enomics (pymatgen): A robust, open-source
  {P}ython library for materials analysis.
\newblock \emph{Computational Materials Science}, 68:\penalty0 314--319, 2013.
\newblock \doi{10.1016/j.commatsci.2012.10.028}.

\bibitem[Packer et~al.(2024)Packer, Wooders, Lin, Fang, Patil, Stoica, and
  Gonzalez]{Packer2024memgpt}
Packer, Charles, Wooders, Sarah, Lin, Kevin, Fang, Vivian, Patil, Shishir~G.,
  Stoica, Ion, and Gonzalez, Joseph~E.
\newblock {MemGPT}: Towards {LLM}s as operating systems, 2024.
\newblock URL \url{http://arxiv.org/abs/2310.08560}.

\bibitem[Paruch and Guyonnet(2013)]{Paruch2013}
Paruch, Patrycja and Guyonnet, Jill.
\newblock Nanoscale studies of ferroelectric domain walls as pinned elastic
  interfaces.
\newblock \emph{Comptes Rendus Physique}, 14\penalty0 (8):\penalty0 667--684,
  2013.
\newblock \doi{10.1016/j.crhy.2013.08.004}.

\bibitem[Qiao et~al.(2024)Qiao, Li, Zhang, He, Kang, Zhang, Yang, Dong, Zhang,
  Wang, Ma, Zhao, Qin, Qin, Du, Xu, Lin, Rajmohan, and
  Zhang]{Qiao2024taskweaver}
Qiao, Bo, Li, Liqun, Zhang, Xu, He, Shilin, Kang, Yu, Zhang, Chaoyun, Yang,
  Fangkai, Dong, Hang, Zhang, Jue, Wang, Lu, Ma, Minghua, Zhao, Pu, Qin, Si,
  Qin, Xiaoting, Du, Chao, Xu, Yong, Lin, Qingwei, Rajmohan, Saravan, and
  Zhang, Dongmei.
\newblock {TaskWeaver}: A code-first agent framework, 2024.
\newblock URL \url{http://arxiv.org/abs/2311.17541}.

\bibitem[Rein et~al.(2024)Rein, Hou, Stickland, Petty, Pang, Dirani, Michael,
  and Bowman]{Rein2024}
Rein, David, Hou, Betty~Li, Stickland, Asa~Cooper, Petty, Jackson, Pang,
  Richard~Yuanzhe, Dirani, Julien, Michael, Julian, and Bowman, Samuel~R.
\newblock {GPQA}: A graduate-level {Google}-proof {Q\&A} benchmark.
\newblock \emph{Proc. COLM}, 2024.

\bibitem[Rosen et~al.(2024)Rosen, Gallant, George, Riebesell, Sahasrabuddhe,
  Shen, Wen, Evans, Petretto, Waroquiers, Rignanese, Persson, Jain, and
  Ganose]{Rosen2024}
Rosen, Andrew~S., Gallant, Max, George, Janine, Riebesell, Janosh,
  Sahasrabuddhe, Hrushikesh, Shen, Jimmy-Xuan, Wen, Mingjian, Evans,
  Matthew~L., Petretto, Guido, Waroquiers, David, Rignanese, Gian-Marco,
  Persson, Kristin~A., Jain, Anubhav, and Ganose, Alex~M.
\newblock Jobflow: Computational workflows made simple.
\newblock \emph{Journal of Open Source Software}, 9\penalty0 (93):\penalty0
  5995, 2024.
\newblock \doi{10.21105/joss.05995}.

\bibitem[Shinn et~al.(2023)Shinn, Cassano, Berman, Gopinath, Narasimhan, and
  Yao]{Shinn2023}
Shinn, Noah, Cassano, Federico, Berman, Edward, Gopinath, Ashwin, Narasimhan,
  Karthik, and Yao, Shunyu.
\newblock Reflexion: Language agents with verbal reinforcement learning, 2023.
\newblock URL \url{http://arxiv.org/abs/2303.11366}.
\newblock NeurIPS 2023.

\bibitem[Sumers et~al.(2024)Sumers, Yao, Narasimhan, and
  Griffiths]{Sumers2024coala}
Sumers, Theodore~R., Yao, Shunyu, Narasimhan, Karthik, and Griffiths, Thomas~L.
\newblock Cognitive architectures for language agents, 2024.
\newblock URL \url{http://arxiv.org/abs/2309.02427}.

\bibitem[Vriza et~al.(2026)Vriza, Kornu, Koneru, Chan, and
  Sankaranarayanan]{Vriza2026}
Vriza, Aikaterini, Kornu, Uma, Koneru, Aditya, Chan, Henry, and
  Sankaranarayanan, Subramanian K. R.~S.
\newblock Multi-agentic {AI} framework for end-to-end atomistic simulations.
\newblock \emph{Digital Discovery}, 5\penalty0 (1):\penalty0 440--452, 2026.
\newblock \doi{10.1039/D5DD00435G}.

\bibitem[Wang et~al.(2023)Wang, Xie, Jiang, Mandlekar, Xiao, Zhu, Fan, and
  Anandkumar]{Wang2023voyager}
Wang, Guanzhi, Xie, Yuqi, Jiang, Yunfan, Mandlekar, Ajay, Xiao, Chaowei, Zhu,
  Yuke, Fan, Linxi, and Anandkumar, Anima.
\newblock Voyager: An open-ended embodied agent with large language models,
  2023.
\newblock URL \url{http://arxiv.org/abs/2305.16291}.

\bibitem[Wang et~al.(2018)Wang, Zhang, Han, and E]{Zhang2018dpmd}
Wang, Han, Zhang, Linfeng, Han, Jiequn, and E, Weinan.
\newblock {DeePMD-kit}: A deep learning package for many-body potential energy
  representation and molecular dynamics.
\newblock \emph{Computer Physics Communications}, 228:\penalty0 178--184, 2018.
\newblock \doi{10.1016/j.cpc.2018.03.016}.

\bibitem[Wang et~al.(2024)Wang, Chen, Yuan, Zhang, Li, Peng, and
  Ji]{Wang2024codeact}
Wang, Xingyao, Chen, Yangyi, Yuan, Lifan, Zhang, Yizhe, Li, Yunzhu, Peng, Hao,
  and Ji, Heng.
\newblock Executable code actions elicit better {LLM} agents, 2024.
\newblock URL \url{http://arxiv.org/abs/2402.01030}.
\newblock ICML 2024.

\bibitem[Xia et~al.(2025)Xia, Ma, Zheng, Zhang, Li, Su, Hu, Zhang, Gong,
  Ouyang, Bai, Zhou, and Su]{Xia2025vasp}
Xia, Zeyu, Ma, Jinzhe, Zheng, Congjie, Zhang, Shufei, Li, Yuqiang, Su, Hang,
  Hu, P., Zhang, Changshui, Gong, Xingao, Ouyang, Wanli, Bai, Lei, Zhou,
  Dongzhan, and Su, Mao.
\newblock An agentic framework for autonomous materials computation, 2025.
\newblock URL \url{http://arxiv.org/abs/2512.19458}.
\newblock arXiv:2512.19458.

\bibitem[Xiao et~al.(2024)Xiao, Tian, Chen, Han, and
  Lewis]{Xiao2024streamingllm}
Xiao, Guangxuan, Tian, Yuandong, Chen, Beidi, Han, Song, and Lewis, Mike.
\newblock Efficient streaming language models with attention sinks.
\newblock \emph{Proc. ICLR}, 2024.

\bibitem[Yao et~al.(2023)Yao, Zhao, Yu, Du, Shafran, Narasimhan, and
  Cao]{Yao2023}
Yao, Shunyu, Zhao, Jeffrey, Yu, Dian, Du, Nan, Shafran, Izhak, Narasimhan,
  Karthik~R., and Cao, Yuan.
\newblock {ReAct}: Synergizing reasoning and acting in language models.
\newblock In \emph{Proc. ICLR}, 2023.
\newblock URL \url{https://openreview.net/forum?id=WE_vluYUL-X}.

\bibitem[Zhang et~al.(2025{\natexlab{a}})Zhang, Li, Xu, Jin, Wu, and
  Li]{Zhang2025topo}
Zhang, Baohua, Li, Xin, Xu, Huangchao, Jin, Zhong, Wu, Quansheng, and Li, Ce.
\newblock {TopoMAS}: Large language model driven topological materials
  multiagent system, 2025{\natexlab{a}}.
\newblock URL \url{http://arxiv.org/abs/2507.04053}.
\newblock arXiv:2507.04053.

\bibitem[Zhang et~al.(2020)]{Zhang2020dpgen}
Zhang, Y. et~al.
\newblock {DP-GEN}: A concurrent learning platform for the generation of
  reliable deep learning based potential energy models.
\newblock \emph{Comput. Phys. Commun.}, 253:\penalty0 107206, 2020.

\bibitem[Zhang et~al.(2025{\natexlab{b}})Zhang, Zhao, Wang, Yang, Wei, and
  Wu]{Zhang2025cast}
Zhang, Yilin, Zhao, Xinran, Wang, Zora~Zhiruo, Yang, Chenyang, Wei, Jiayi, and
  Wu, Tongshuang.
\newblock {cAST}: Enhancing code retrieval-augmented generation with structural
  chunking via abstract syntax tree, 2025{\natexlab{b}}.
\newblock URL \url{http://arxiv.org/abs/2506.15655}.

\bibitem[Zheng et~al.(2025)Zheng, Florit, Jin, Wu, Li, Nandiwale, Salazar,
  Mustakis, Green, and Jensen]{Zheng2025}
Zheng, Zhiling, Florit, Federico, Jin, Brooke, Wu, Haoyang, Li, Shih-Cheng,
  Nandiwale, Kakasaheb~Y., Salazar, Chase~A., Mustakis, Jason~G., Green,
  William~H., and Jensen, Klavs~F.
\newblock Integrating machine learning and large language models to advance
  exploration of electrochemical reactions.
\newblock \emph{Angewandte Chemie International Edition}, 64\penalty0
  (6):\penalty0 e202418074, 2025.
\newblock \doi{10.1002/anie.202418074}.

\bibitem[Zou et~al.(2025)Zou, Cheng, Aldossary, Bai, Leong,
  Campos-Gonzalez-Angulo, Choi, Ser, Tom, Wang, Zhang, Yakavets, Hao,
  Crebolder, Bernales, and Aspuru-Guzik]{Zou2025}
Zou, Yunheng, Cheng, Austin~H., Aldossary, Abdulrahman, Bai, Jiaru, Leong,
  Shi~Xuan, Campos-Gonzalez-Angulo, Jorge~Arturo, Choi, Changhyeok, Ser,
  Cher~Tian, Tom, Gary, Wang, Andrew, Zhang, Zijian, Yakavets, Ilya, Hao, Han,
  Crebolder, Chris, Bernales, Varinia, and Aspuru-Guzik, Al\'{a}n.
\newblock El {A}gente: An autonomous agent for quantum chemistry.
\newblock \emph{Matter}, 8\penalty0 (7), 2025.
\newblock \doi{10.1016/j.matt.2025.102263}.

\end{thebibliography}

\end{document}